# Natural Deduction Derivations as Directed Acyclic Proof Graphs

Jonathan Chan and Aaron Stump
Boston College

June – July 2026

## 1 Introduction

In conventional natural deduction, a proof of a proposition is a derivation tree constructed out of inference rules with some number of premises and a conclusion judgement. For instance, the following are the usual rules for conjunctions and disjunctions.

$$\frac{\Gamma \vdash A \quad \Gamma \vdash B}{\Gamma \vdash A \land B} \land i \qquad \frac{\Gamma \vdash A_1 \land A_2}{\Gamma \vdash A_k} \land e_k$$

$$\frac{\Gamma \vdash A_k}{\Gamma \vdash A_1 \lor A_2} \lor i_k \qquad \frac{\Gamma \vdash A \lor B \quad \Gamma, A \vdash C \quad \Gamma, B \vdash C}{\Gamma \vdash C} \lor e$$

In contrast to the other rules, the disjunction elimination rule has an asymmetry to it because it involves some formula $C$ unrelated to either of the subformulae involved in the disjunction. Aside from its aesthetics, involvement of a tertiary propositional also complicates proof simplification. Consider taking the first conjunct resulting from a disjunction elimination.

$$\frac{\dfrac{\vdash C \lor D \quad C \vdash A \land B \quad D \vdash A \land B}{\vdash A \land B} \lor e}{\vdash B} \land e_1$$

From a computational point of view, this is taking the first projection of a pair of $A$ and $B$ returned by a conditional expression branching on $C$ or $D$. Therefore, it ought to be morally equivalent to take the projections in each of the branches instead.

$$\frac{C \lor D \quad \dfrac{C \vdash A \land B}{C \vdash B} \land e_1 \quad \dfrac{D \vdash A \land B}{D \vdash B} \land e_1}{\vdash B} \lor e$$

Doing this *commuting conversion* and pushing the projection into the branches is desirable, since it could lead to further proof simplification if the branches were to be conjunction literals. In terms of the simply typed lambda calculus, such a commutation corresponds to the following syntactic transformation:

$$(\text{case } cd \text{ of inl } c \Rightarrow f \mid \text{inr } d \Rightarrow g)\ a$$
$$\text{case } cd \text{ of inl } c \Rightarrow (f\ a) \mid \text{inr } d \Rightarrow (g\ a)$$

Sadly, these two derivations aren't $\beta$-equivalent, and commuting conversions are typically treated as compiler passes rather than *bona fide* $\beta$-reductions, since doing them naïvely may result in excessive code duplication and requires special handling.

To resolve the asymmetry of disjunction elimination, Girard (1989, Chapter 10.1) suggests a rule with multiple conclusions, but claims that "we have no way of bringing them back together".

$$\frac{\vdash A \vee B}{\vdash A \quad \vdash B}\vee e'$$

In this project, we pursue his proposal using the key insight that proof derivations need not be trees, but may be *directed acyclic graphs* (DAGs) in general. Derivation rules are represented as nodes, premises as incoming edges, and conclusions as outgoing edges. Treating derivations as DAGs permits using the alternate rule by ensuring that the subderivations pointed at by each conclusion eventually join up at a new concluding node. Commuting conversions can then be formulated in terms of interactions with this join node, which also acts as a join point from a computational perspective.

The primary goal of this project is to design a proof graph language with reduction rules such that proof simplification retains its concluding judgements, *i.e.* type preservation. These reduction rules should also encompass the various possible commuting conversions.

## 2 Typing Rules as Labelled Graph Nodes

A typing rule consists of a rule name and the typing judgements that make up its premises and conclusions. We represent a rule graphically as a named node with incoming edges labelled by its premises and outgoing edges labelled by its conclusions. A complete derivation is constructed by connecting rule nodes along matching typing edges. We assume to not be dealing with cyclic proofs, and so our derivations are directed acyclic graphs. The rules for some basic connectives we keep unmodified are given in Figure 1.

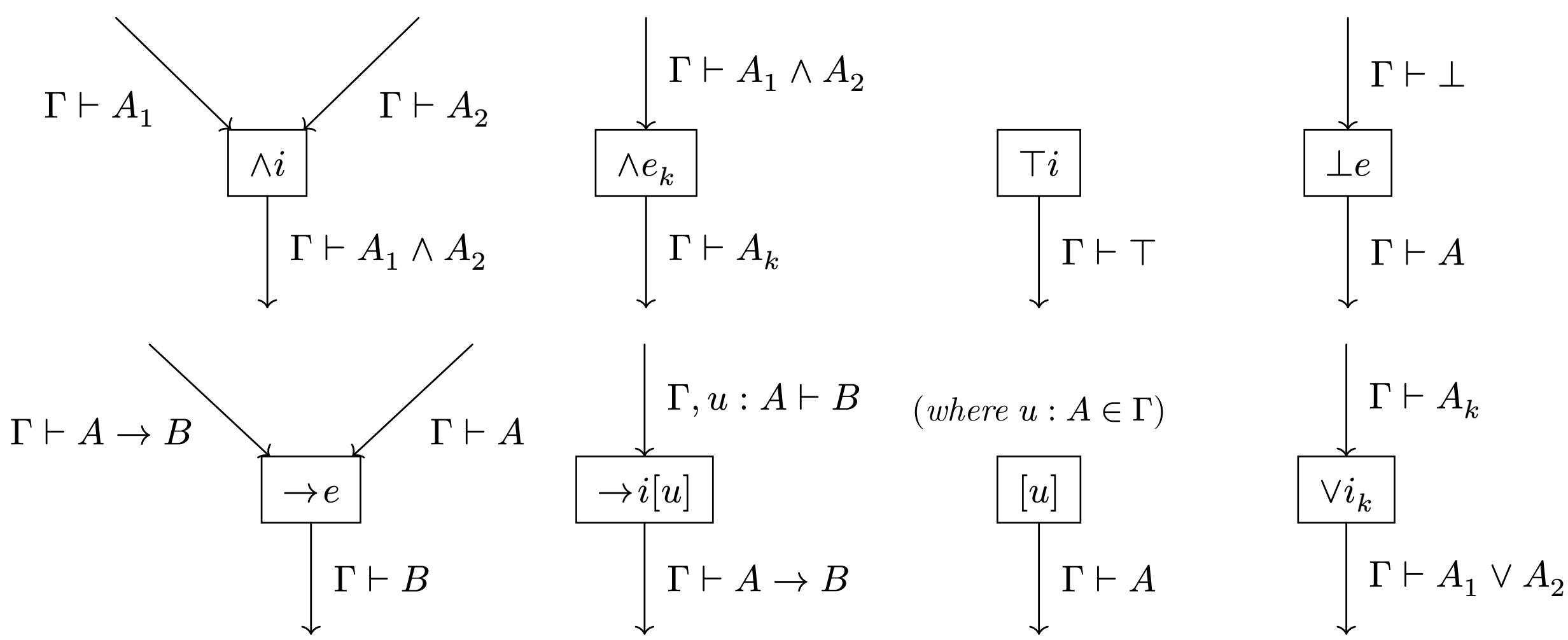

Figure 1: Typing rules — conjunction, true, false, implication, disjunction introduction

The disjunction elimination rule proposed by Girard would then have multiple outgoing conclusion edges. However, these edges do not represent independent conclusions—it would be wrong to conclude $\Gamma \vdash A$ out of $\Gamma \vdash A \vee B$.

Instead, we introduce a new *join node* $\oplus$, which joins up the two branches of deductions out of eliminating a disjunction. To track which branch of which disjunction a path of deduction is in, we use a *control context* $\Theta$ containing declarations $x : k$. A disjunction elimination labelled by $x$ concludes with judgements whose control contexts are extended by $x : 1$ and $x : 2$ respectively, and a join node labelled by $x$ may only join two premises from different branches. Just as the usual Barendregt convention assumes each variable of $\to i[u]$ to be distinct, we also assume each variable of $\vee e[x]$ to be distinct. Finally, a *terminal node* $\bullet$ completes a proof derivation by requiring an empty control context, representing that all paths have been handled. These new nodes and rules are given in Figure 2; we consider the latter two to be *structural*, as they aren't associated with a particular logical connective.

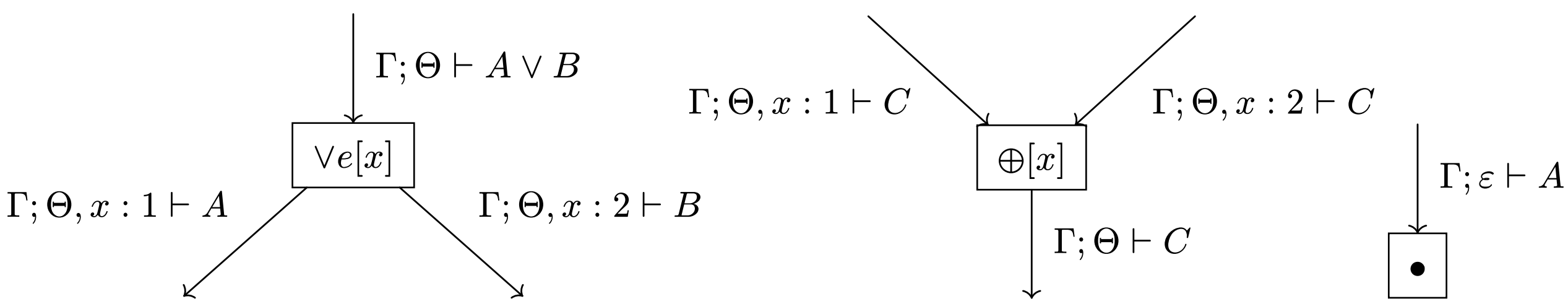

Figure 2: Typing rules — disjunction elimination, join, terminal

For now, the rules in Figure 1 are updated by inserting an arbitrary control context $\Theta$ everywhere. However, doing so will permit joining nonexistent branches, such as in the example of Figure 3.

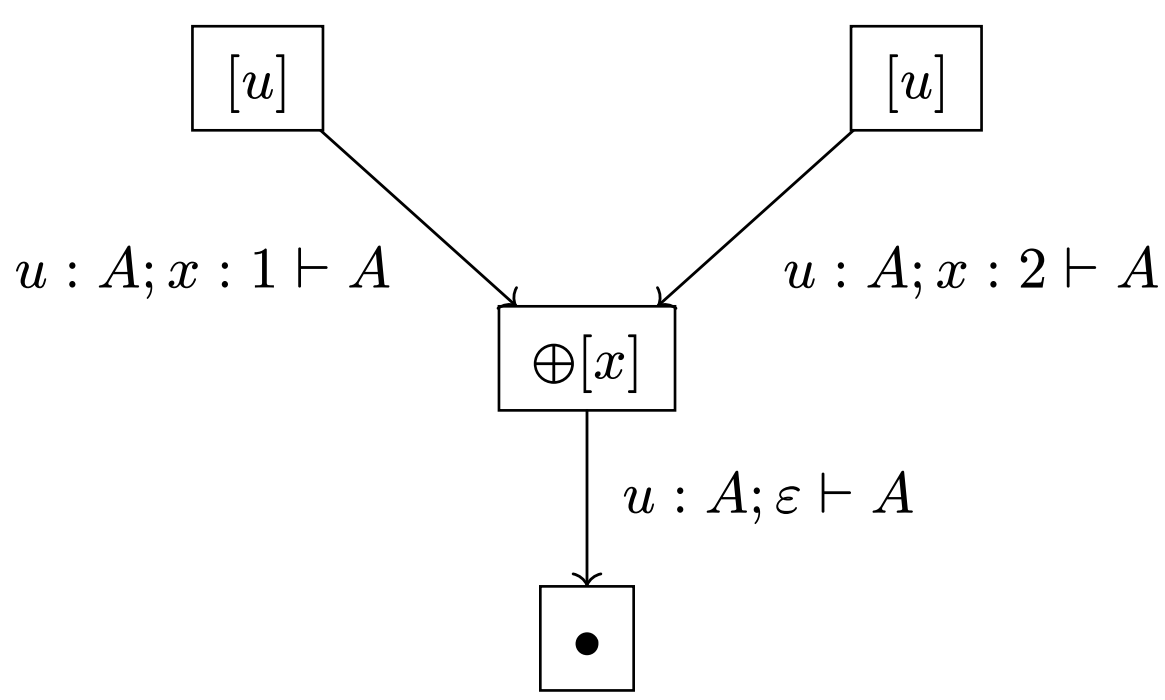

Figure 3: A pathological example

There are two options for ruling out such pathological derivations: requiring empty control contexts in leaf nodes $[u]$ and $\top i$, or requiring that there exists a $\vee e[x]$ node whenever there is a $\oplus[x]$ node somewhere in the derivation. The former would require alternate mechanisms for weakening the control context when needed in sensible ways that rule out the above pathology. The latter may be considered a *global* restriction, since it relies on examining the entire proof graph (which may be disconnected!), in contrast to the former as a *local* restriction that propagates via typing. Later in Section 3.2, we revisit looking at proof graphs locally *vs.* globally in the context of $\beta$-reductions to pick the corresponding option.

As a simple example of a well-typed proof graph nontrivially involving disjunctions and joins, the below is a closed proof of commutativity of disjunction, $A \vee B \to B \vee A$. It would be equivalent to the lambda term $\lambda x.\, \text{case } x \text{ of inl } a \Rightarrow \text{inr } a \mid \text{inr } b \Rightarrow \text{inl } a$.

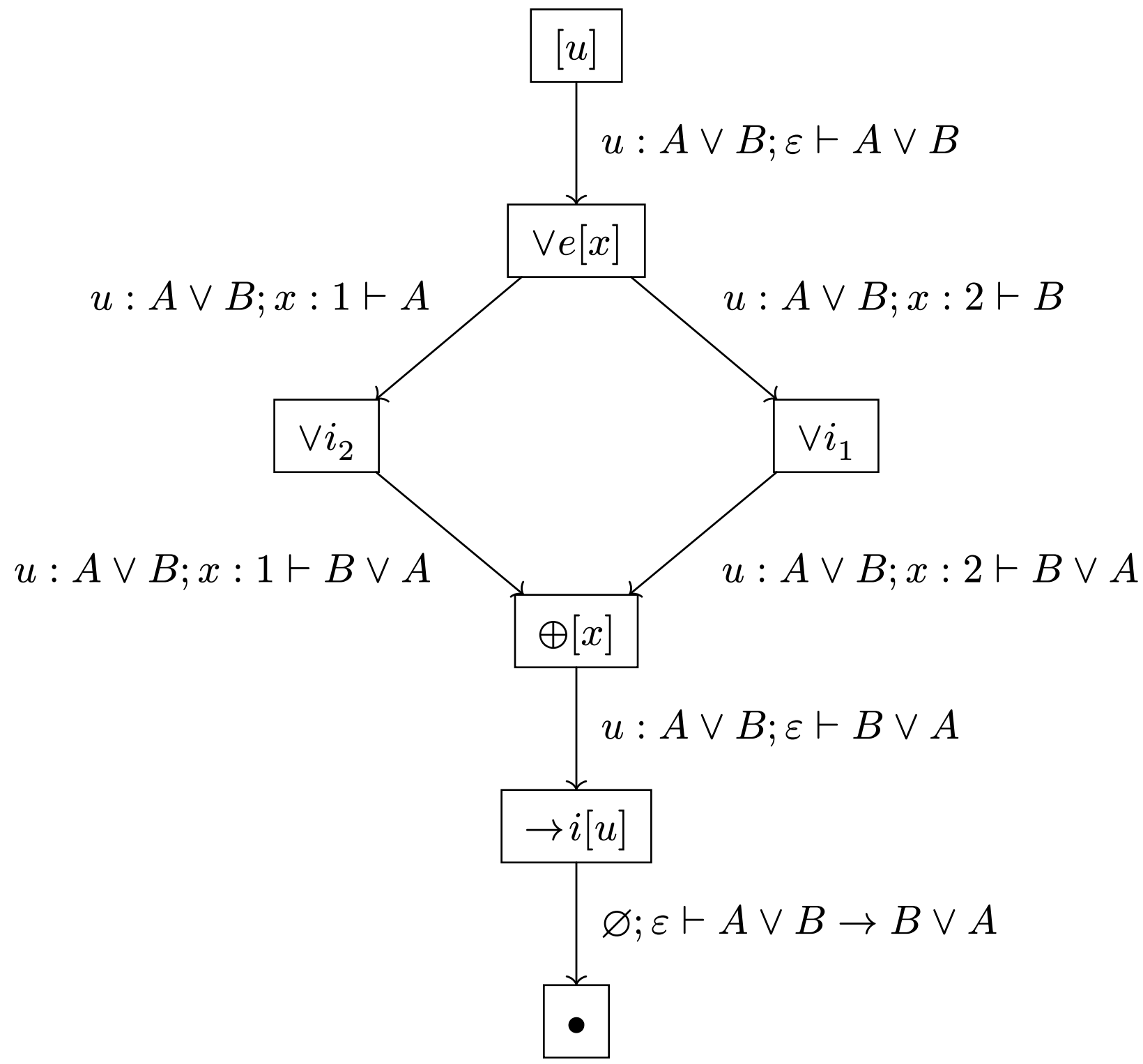

# 3 Computation on Proof Graphs

From the perspective of proof derivations, a computation is a transformation from one derivation to another. In tree-based natural deduction, a reduction rule identifies a pattern of premises, rules, and a conclusion as a subtree, and transforms it into a different subtree with the same premises and conclusion. Correspondingly, in a proof graph, a reduction rule identifies a subgraph of nodes and their incoming and outgoing edges, and transforms it into a different subgraph with the same incoming and outgoing edges (although the edges internal to the subgraph may change). We present them in small-step form, transforming a graph one piece at a time.

While laying out the various reduction rules for proof graphs, we will find the need to further extend our language of nodes with more structural rules even before considering how disjunctions and join behave. Two of these arise from looking at $\beta$-reduction of conjunctions and implications (Section 3.1). For $\beta$-reduction of disjunctions, we consider either a local reduction style (Section 3.2.1), which introduces another three structural nodes, or a global reduction style (Section 3.2.2), which requires only one new node at the expense of compositionality.

## 3.1 Holes and Forks

The first reduction rule we consider is the reduction of conjunction elimination. From a computational perspective, a conjunction is a pair, its elimination a projection, and the reduction rule projecting the corresponding component. The left-hand side of the reduction rule in Figure 4 identifies a subgraph with a conjunction introduction as a premise to an elimination, with two conjuncts as incoming edges to the subgraph and the projection as the outgoing edge.

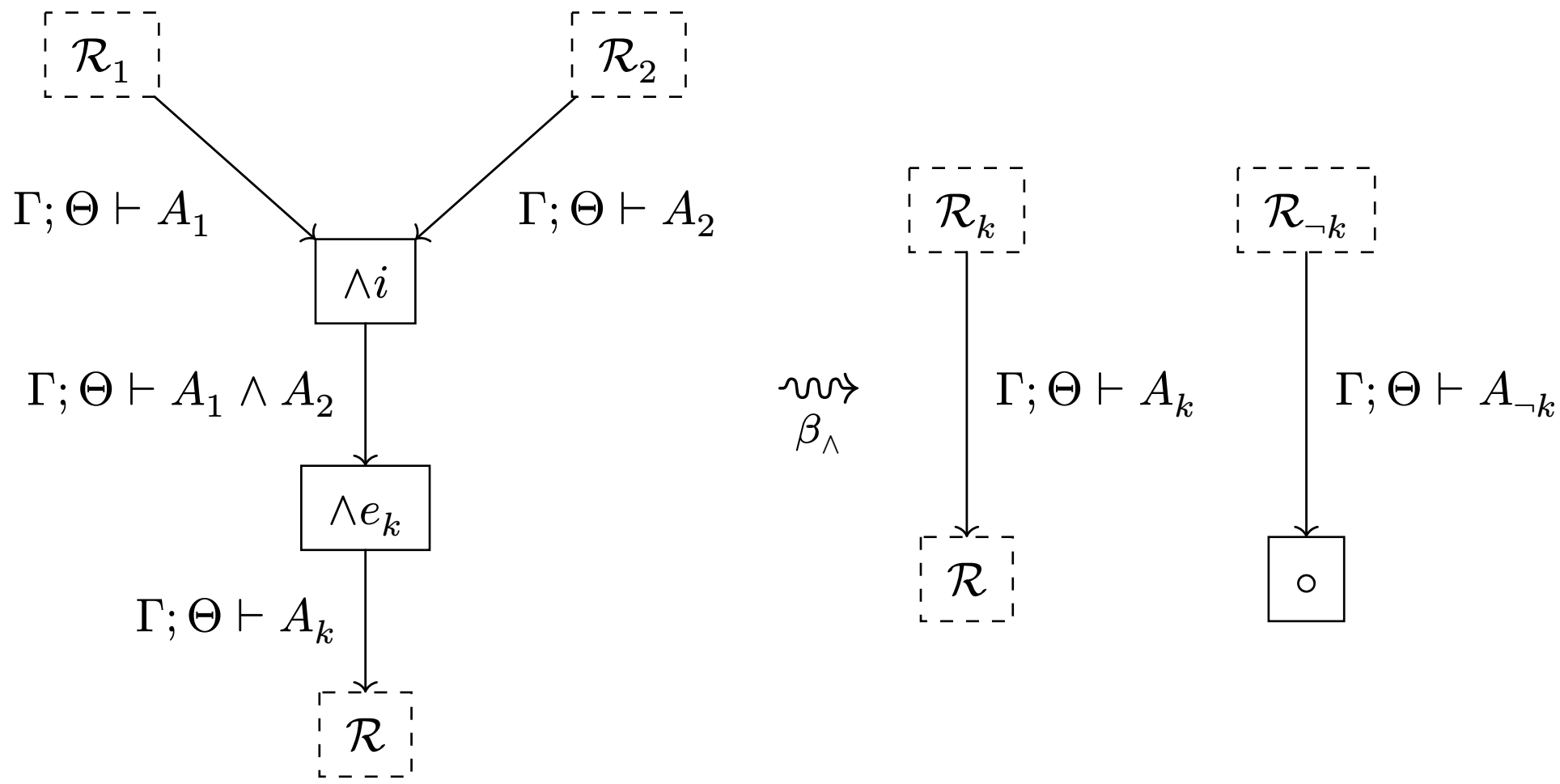

Figure 4: Reduction rule for conjunction elimination

To distinguish where those edges come from and go to, we use dashed nodes to represent rules outside of this subgraph in the rest of the whole graph. While all edges of nodes inside the subgraph are illustrated, the nodes outside of the subgraph may have other incoming and outgoing edges irrelevant to the reduction.

As expected, the right-hand side links the $k$th conjunct to the outgoing projection. But deleting the introduction and elimination nodes leaves a dangling incoming edge from the other conjunct! Since the other conjunct isn't used, we link it to a new *hole node* $\circ$ representing a subgraph that may now be discarded. A proof graph ending in a hole does *not* mean we have a proof of the hole's typing judgement: its control context need not be empty, and so is not a valid proof, since there may be unhandled branches in the path leading up to the hole.

Next, we consider the reduction of implication elimination. From a computational point of view, an implication is a function, its elimination an application, and the reduction rule substituting for each variable node. In the left-hand side of the reduction rule in Figure 5, $\mathcal{G}$ represents the subgraph that comprises the body of the function introduced by $\to i[u]$. This subgraph consists of all nodes that can reach $\to i[u]$, so that the $n$ variable nodes represent all uses of the variable $u$. (It then has no other incoming edges, but may have other irrelevant outgoing edges.)

If $\mathcal{R}_0$ were the root of a tree, then that tree can be copied wholesale $n$ times and linked to $\mathcal{G}$ in place of each $[u]$, each copy weakened by $\Gamma_1, ..., \Gamma_n$ respectively. Unfortunately, $\mathcal{R}_0$ could be, for instance, a disjunction elimination with another outgoing edge for its second branch (ignoring issues of control context weakening, which we deal with in the next two sections). This prevents it both from being copied directly and from being weakened, it being unclear how that second branch ought to be handled.

Therefore, in the right-hand side of the reduction, we use a new *fork node* $\pitchfork$ that at once multiplicates a conclusion and also weakens each typing context independently. The fork node acts as a delayed substitution, permitting $\mathcal{G}$ to use the function argument $n$ times without fear of being blocked by a disjunction elimination.

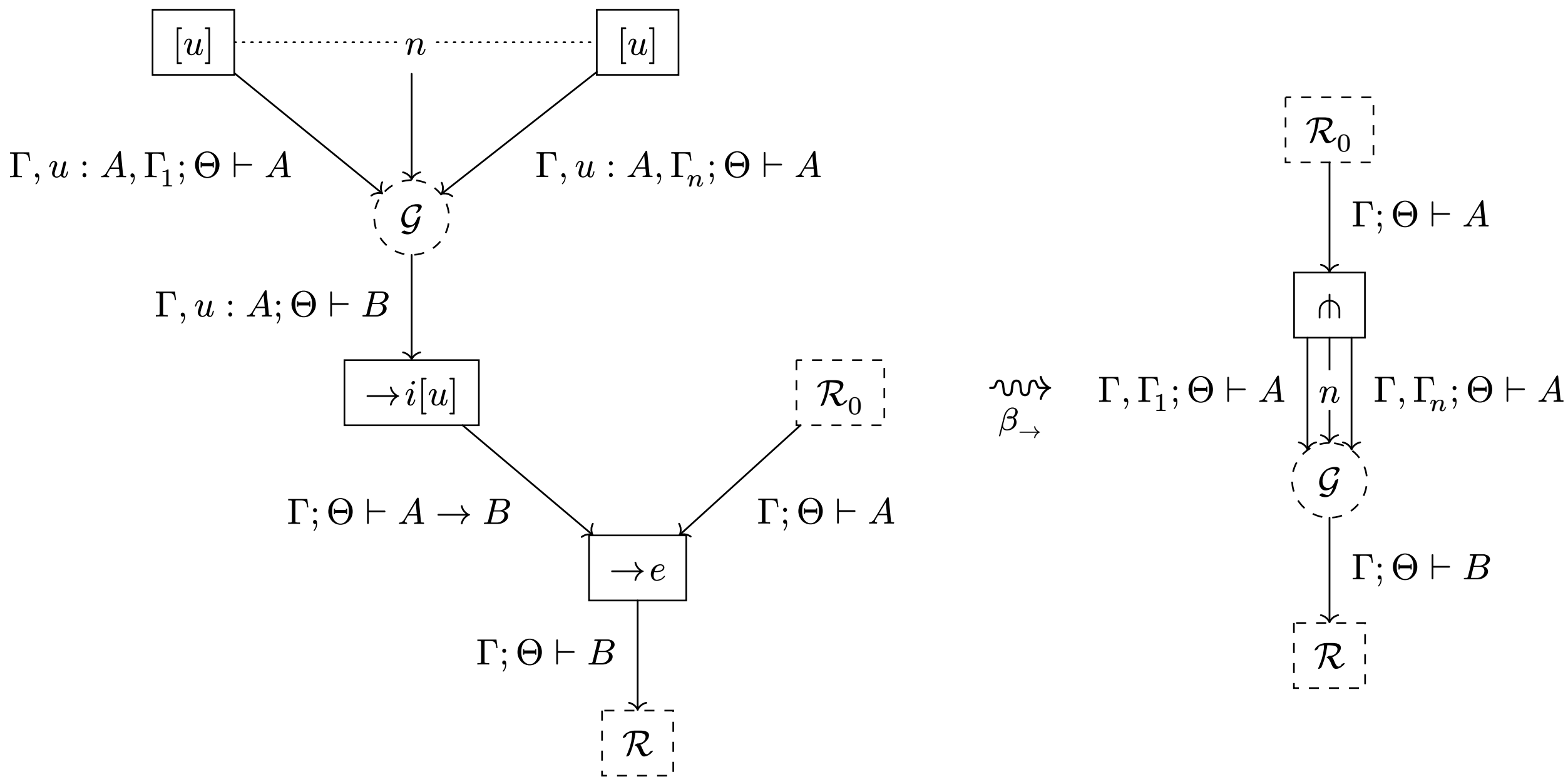

Figure 5: Reduction rule for implication elimination

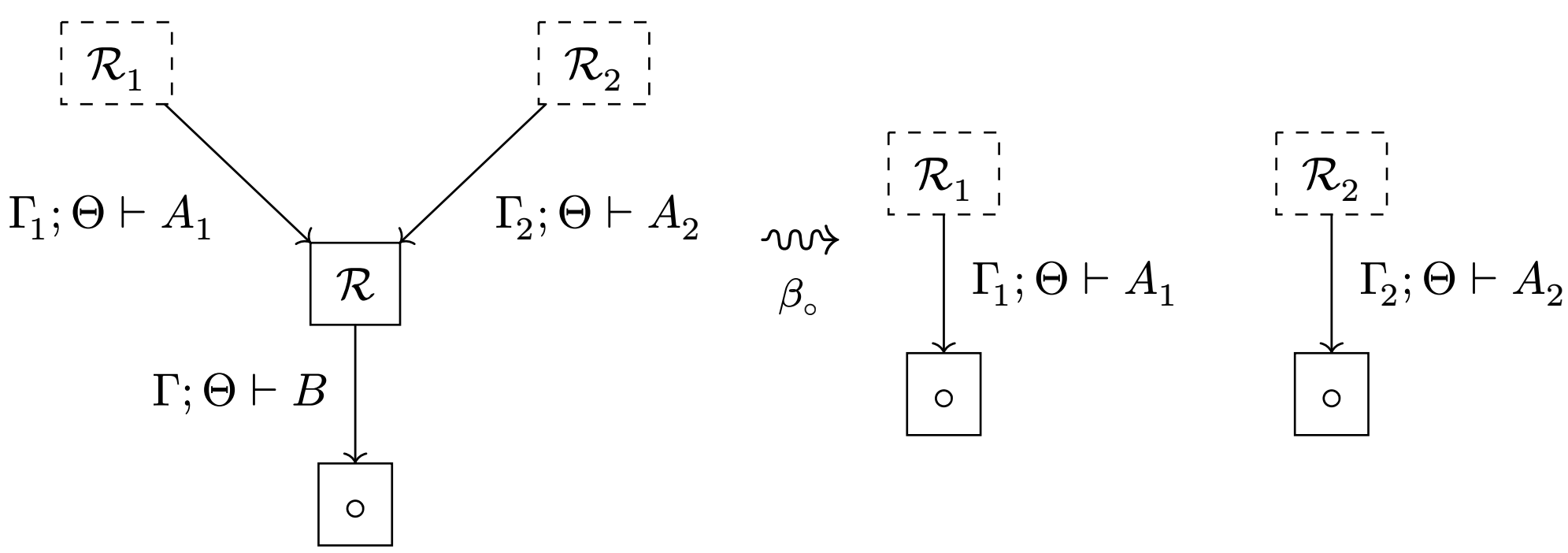

Figure 6: Reduction rule schema for holes

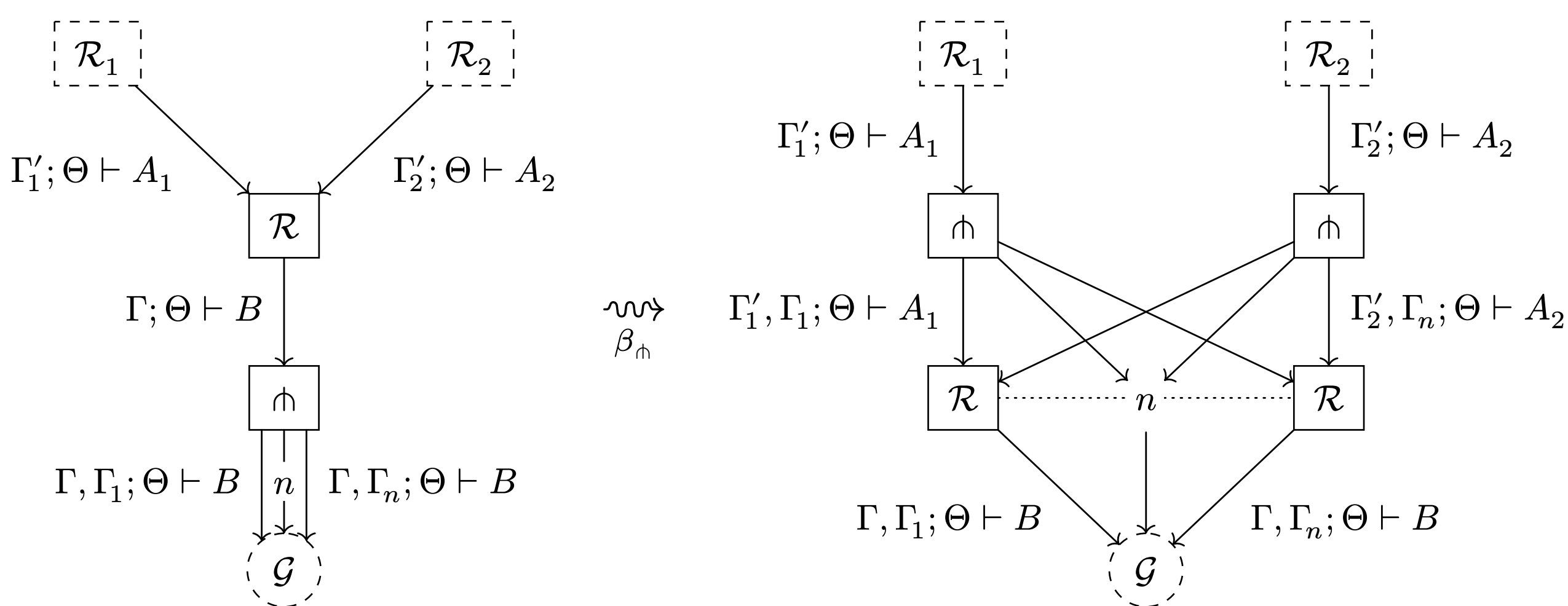

Figure 7: Reduction rule schema for forks

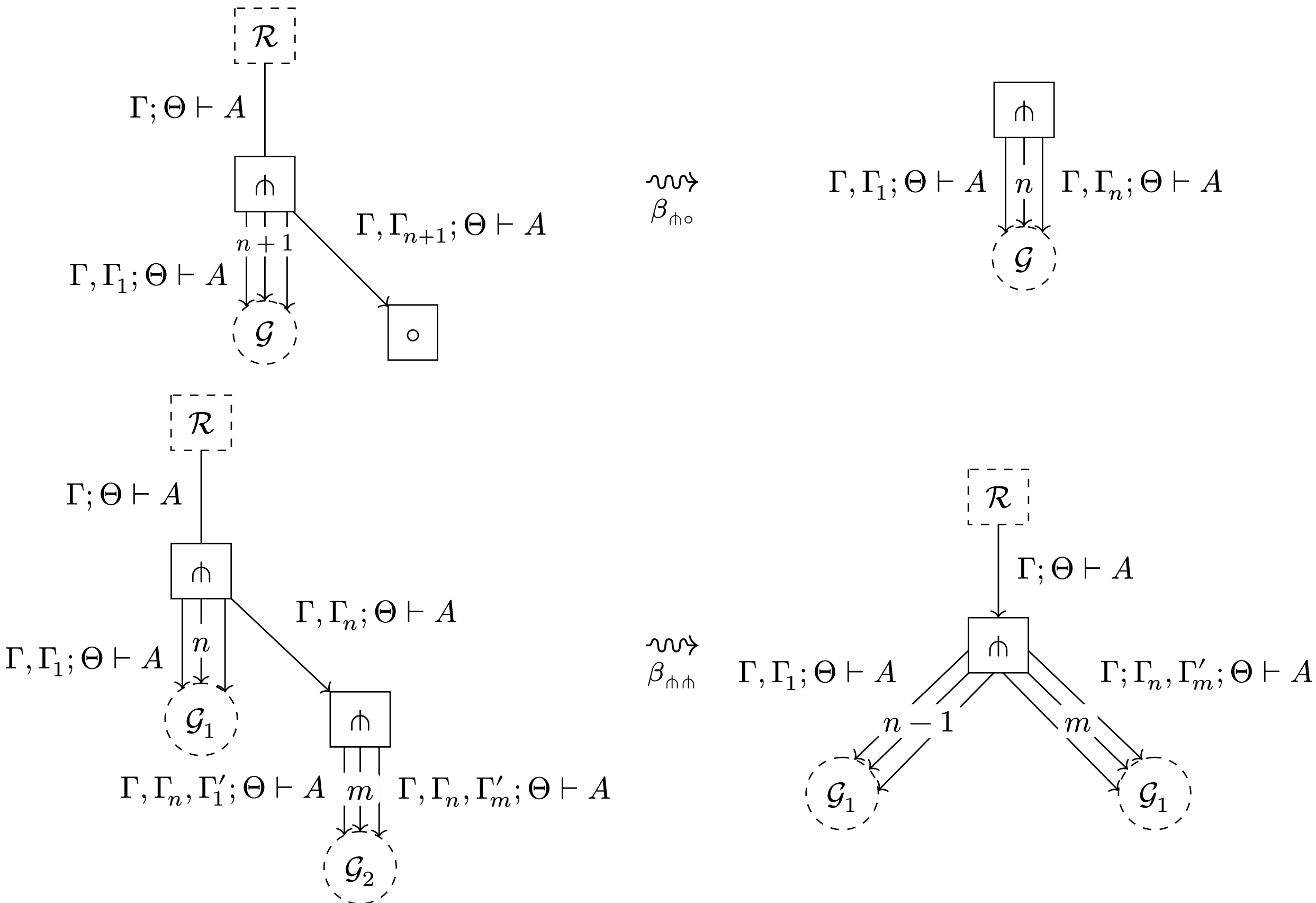

Figure 8: Special reduction rules for forks

The hole and fork nodes themselves have reduction rules: hole nodes swallow single-conclusion nodes, while fork nodes copy single-conclusion nodes. Both are blocked by the double-conclusion disjunction elimination node. The general form of their reduction rules for two-premise nodes are given in Figures 6 and 7. Correspondingly, for one-premise nodes, there is only one hole or fork on the right-hand side. A zero-premise node such as $\top i$ linked to a hole is discarded entirely (*i.e.* reduces to the empty graph), while a zero-premise node linked to a fork with $n$ conclusions is simply multiplicated $n$ times. There are two special cases (Figure 8): when a fork links to a hole, that multiplicate is simply discarded, and when a fork links to another fork, they are combined into one single fork.

With the fork node, it becomes possible to multiplicate the branches of a disjunction elimination $n$ times, then join pairs of branches together using $n$ join nodes. A disjunction elimination with control variable $x$ is analogous to an implication introduction introducing a variable in the typing context, and join nodes with $x$ are analogous to variable nodes that may be used as many times as desired (so long as those uses are well typed).

## 3.2 Reducing Disjunction Elimination

A disjunction elimination of a left or right disjunct reduces to the disjunct linked to the left or right branch, respectively, dropping the other branch. The information of which branch is taken must then be propagated to all join nodes, which reduces by selecting the correct branch.

A local reduction strategy identifies a fixed pattern of nodes to reduce. Therefore, reducing a disjunction is a separate step from reducing a join, since there may be an arbitrary number of nodes between a disjunction elimination node and a join node. Propagating which branch is taken or dropped requires additional structural nodes to carry the information from disjunction to join node step by step.

In contrast, a global reduction strategy looks at the entire proof graph as a whole, including all of its disconnected components. In a single step, it identifies a disjunction to reduce, along with *all* of its join nodes, linking the taken branch and unlinking the dropped branch from both ends. Such reductions aren't contextually closed, since linking the left- and right-hand sides of the reduction to an outer piece of graph may miss reducing join nodes in that piece.

We examine these two approaches separately in the subsections to follow. As a running example, we consider the desired evaluation in Figure 9, which must always reduce from the left-hand side to the right in some number of steps under both strategies. The key feature of this example is that the actual disjunct $\mathcal{R}_0$ is discarded, but which branch is taken dictates which premise of $\mathcal{R}_1, \mathcal{R}_2$ gets linked to the conclusion $\mathcal{R}$.

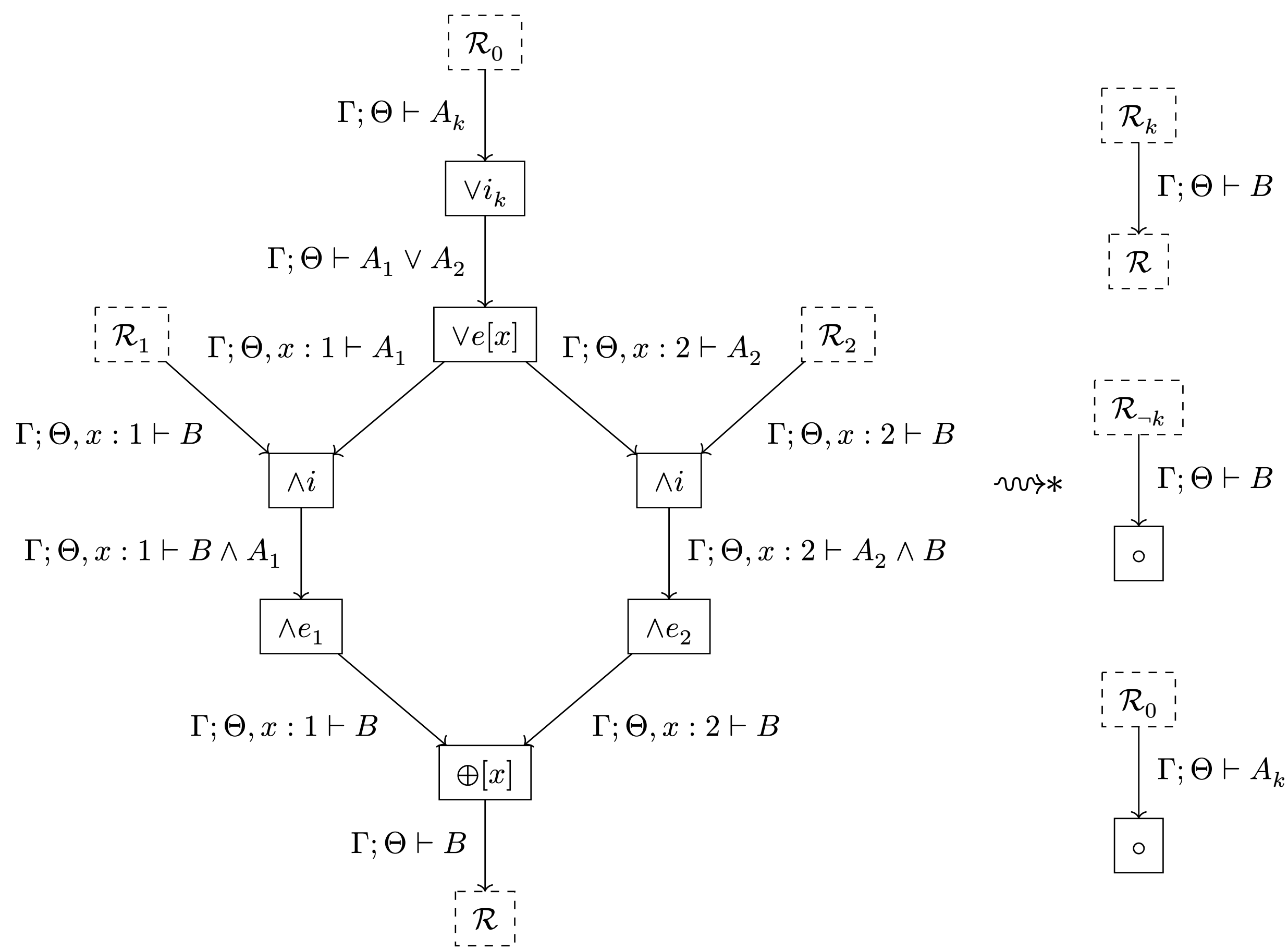

Figure 9: Example evaluation of a disjunction elimination

### 3.2.1 Local Reduction

A reduction occurs when a disjunction introduction is a premise to an elimination, shown on the left-hand side of Figure 10 (ignoring the annotation on the new declaration in the control context for now). Since the $k$th disjunct is introduced, we expect $\mathcal{R}_0$ to be linked to $\mathcal{R}_k$. However, doing so directly would require weakening the control context of $\mathcal{R}_0$ by $x : k$; recall also that to rule out the pathological example in Figure 3 locally, we prohibit implicit control weakening via arbitrary control contexts in leaf nodes.

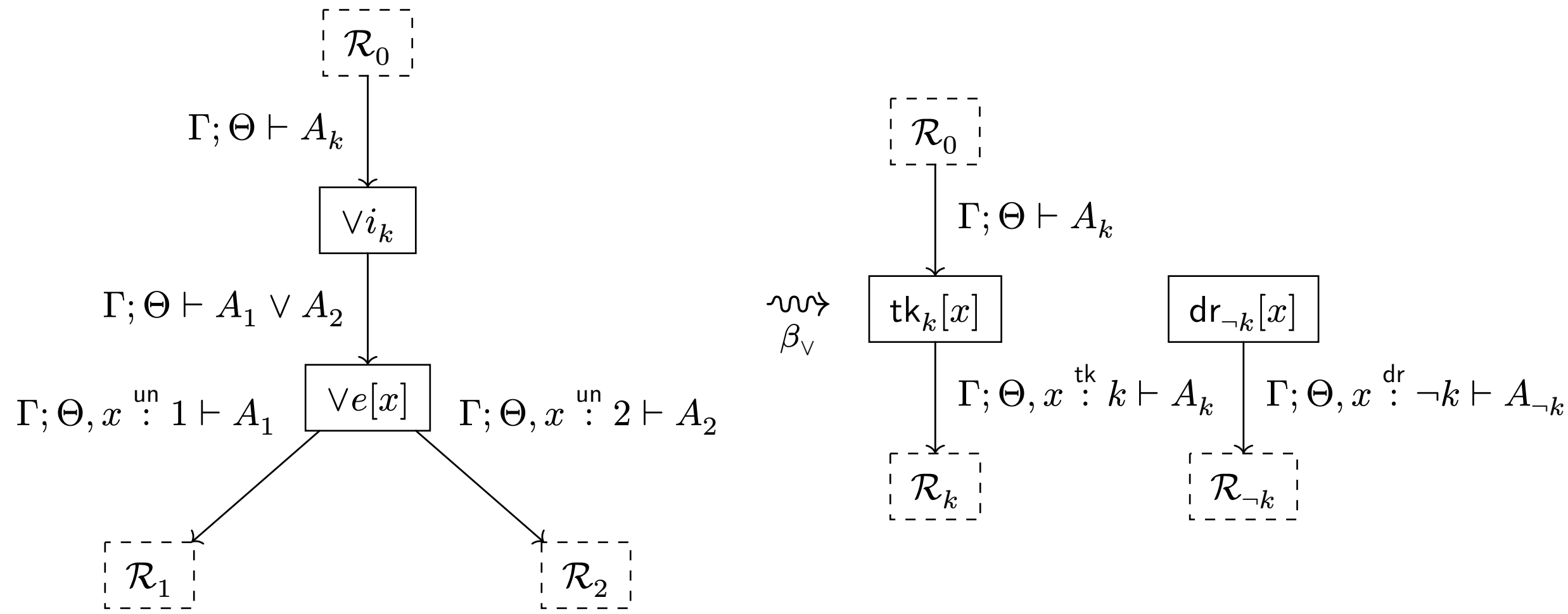

Figure 10: Local reduction rule for disjunction elimination

Therefore, on the right-hand side of the reduction, the disjunct is linked to the $k$th conclusion via an explicit *take node* $\mathsf{tk}$, which indicates that the $k$th branch is taken by carrying the disjunct as a premise and explicitly weakens the control context by $x : k$. Conversely, we link a *drop node* $\mathsf{dr}$ to the branch that isn't taken; it acts as a placeholder that will eventually be consumed by a hole node. While their types don't change, each $\mathsf{un}$ annotation does change to signal that more information was gained.

Unfortunately, by introducing nodes that extend the control context, we have reïntroduced the pathology! Figure 11 gives seemingly well-typed derivations of proofs nonsensically stating that the branches being joined are *both* taken or *both* dropped.

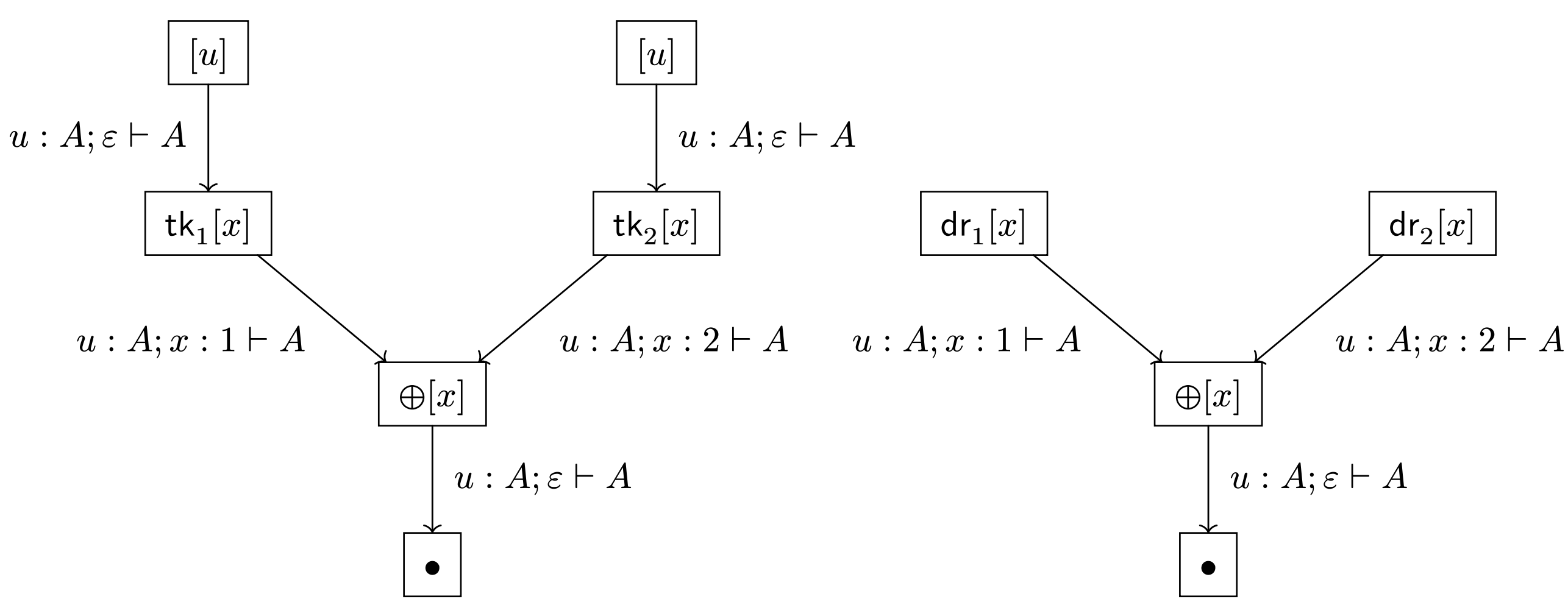

Figure 11: Pathological examples with take and drop

Thankfully, control declarations can only come from three distinct nodes: disjunction elimination nodes, take nodes, and drop nodes. Consequently, a branch must either be known to be taken or dropped, or not yet known. This information can be propagated to join nodes by annotating control declarations with one of three annotations $\delta$: $\mathsf{un}$ for unknown branches, $\mathsf{tk}$ for taken branches, and $\mathsf{dr}$ for dropped branches.

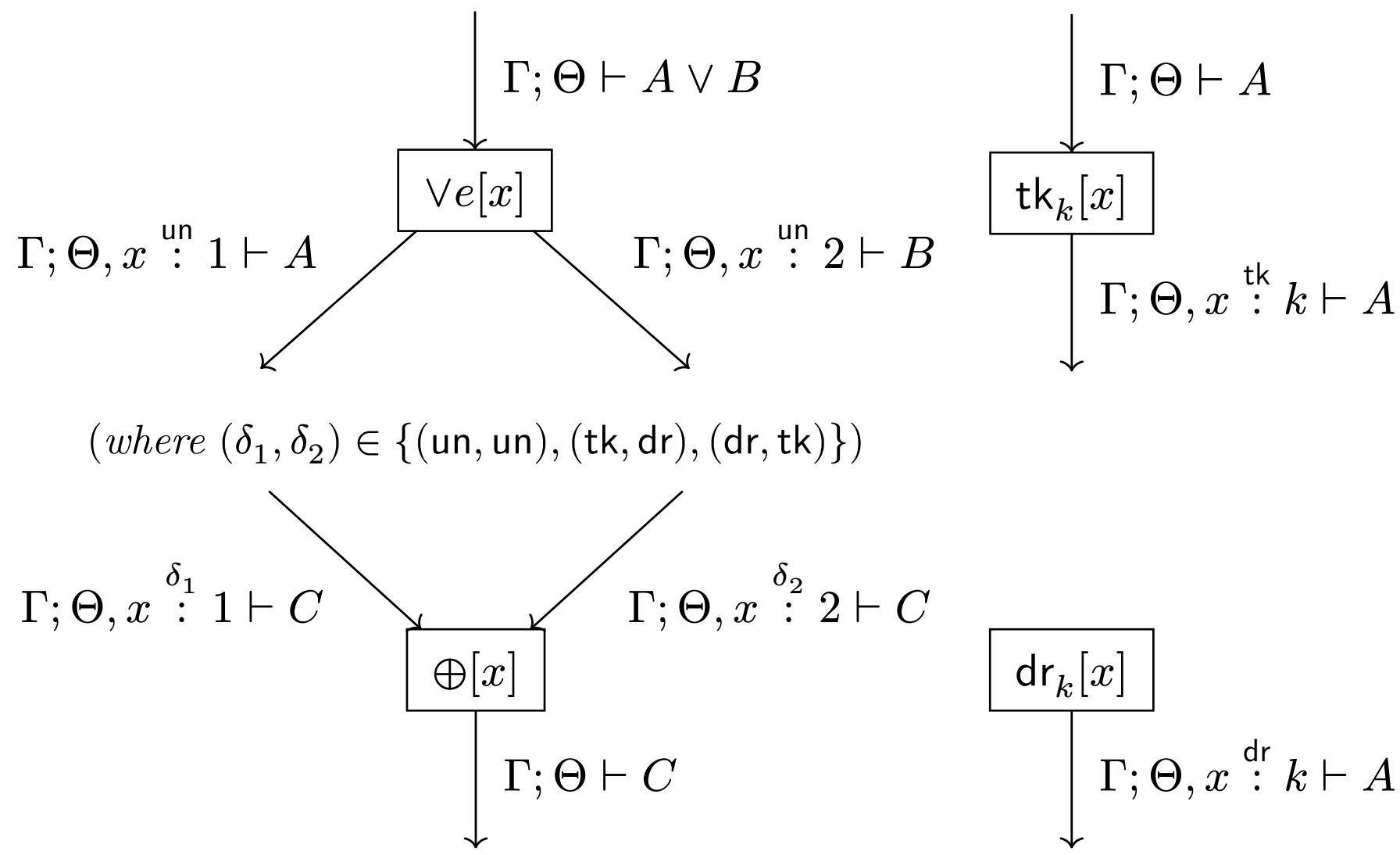

Figure 12: Typing rules — disjunction elimination, join, take, drop

Then pathologies are ruled out by restricting in the typing rule what kinds of branches may be joined. The relevant annotated typing rules are listed in Figure 12. In particular, a join node may only join branches where both are unknown (since if one is known to be taken or dropped, then the other must necessarily also be known), or where exactly one branch is taken and the other dropped. These join conditions make typing the pathologies of Figure 11 impossible.

A join node then reduces when it encounters a take node, linking the taken premise to its conclusion and discarding the dropped branch into a hole. The rule in Figure 13 illustrates taking the first branch, with a dropped first branch and a taken second branch reducing similarly.

Take nodes reach join nodes by themselves reducing when they are premises to other nodes, propagating past those nodes if they aren't join nodes of the same control variable. When all premises are take nodes, they move downwards into the conclusions. Figure 14 shows the general form of this reduction.

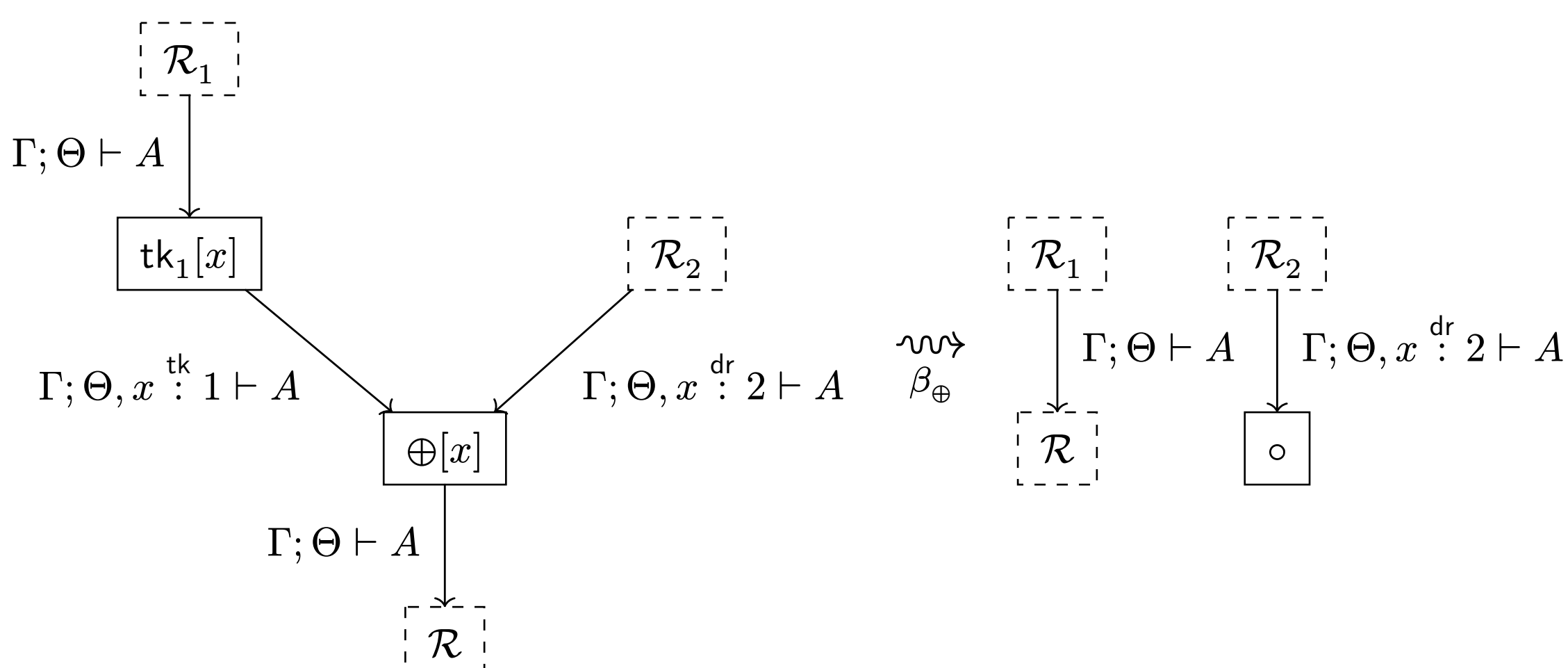

Figure 13: Reduction rule for join (first branch)

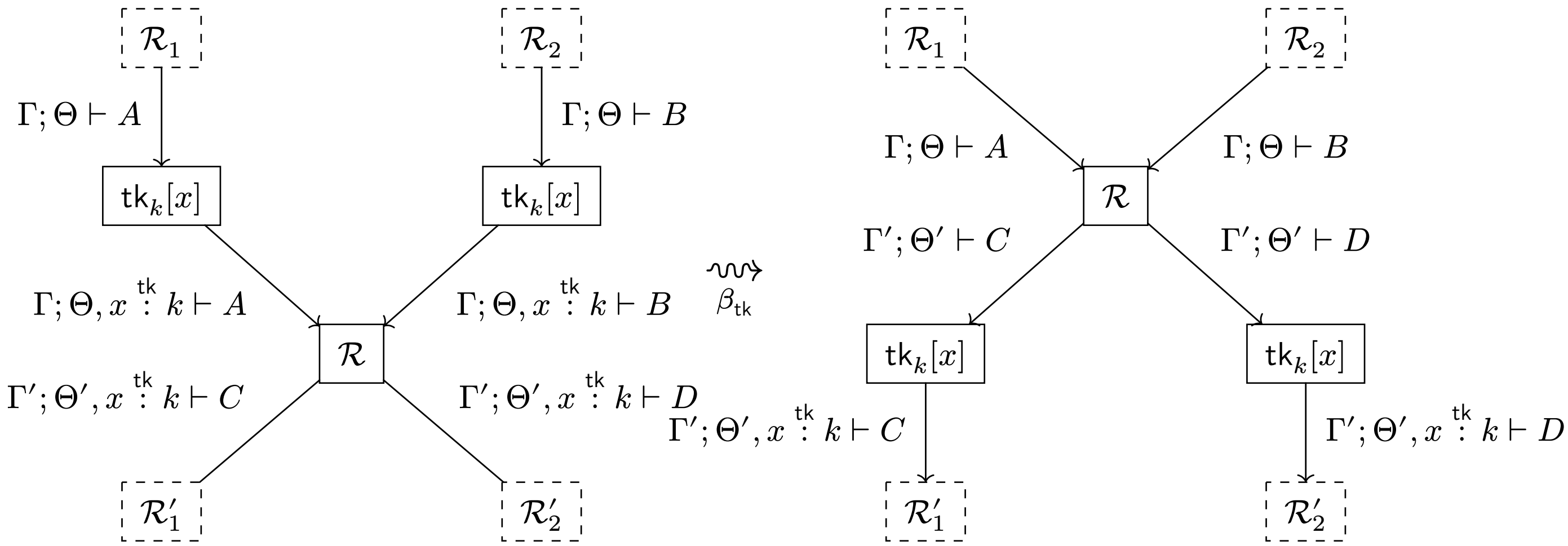

Figure 14: Reduction rule schema for take

Now we are finally ready to revisit the example from Figure 9. In particular, we first look at just the beginning of the first branch, reproduced in Figure 15 with annotations. Since $\mathcal{R}_1$'s outgoing edge has a control context containing $x \overset{\mathsf{un}}{:} k$, there must have been an explicit weakening in its premises.

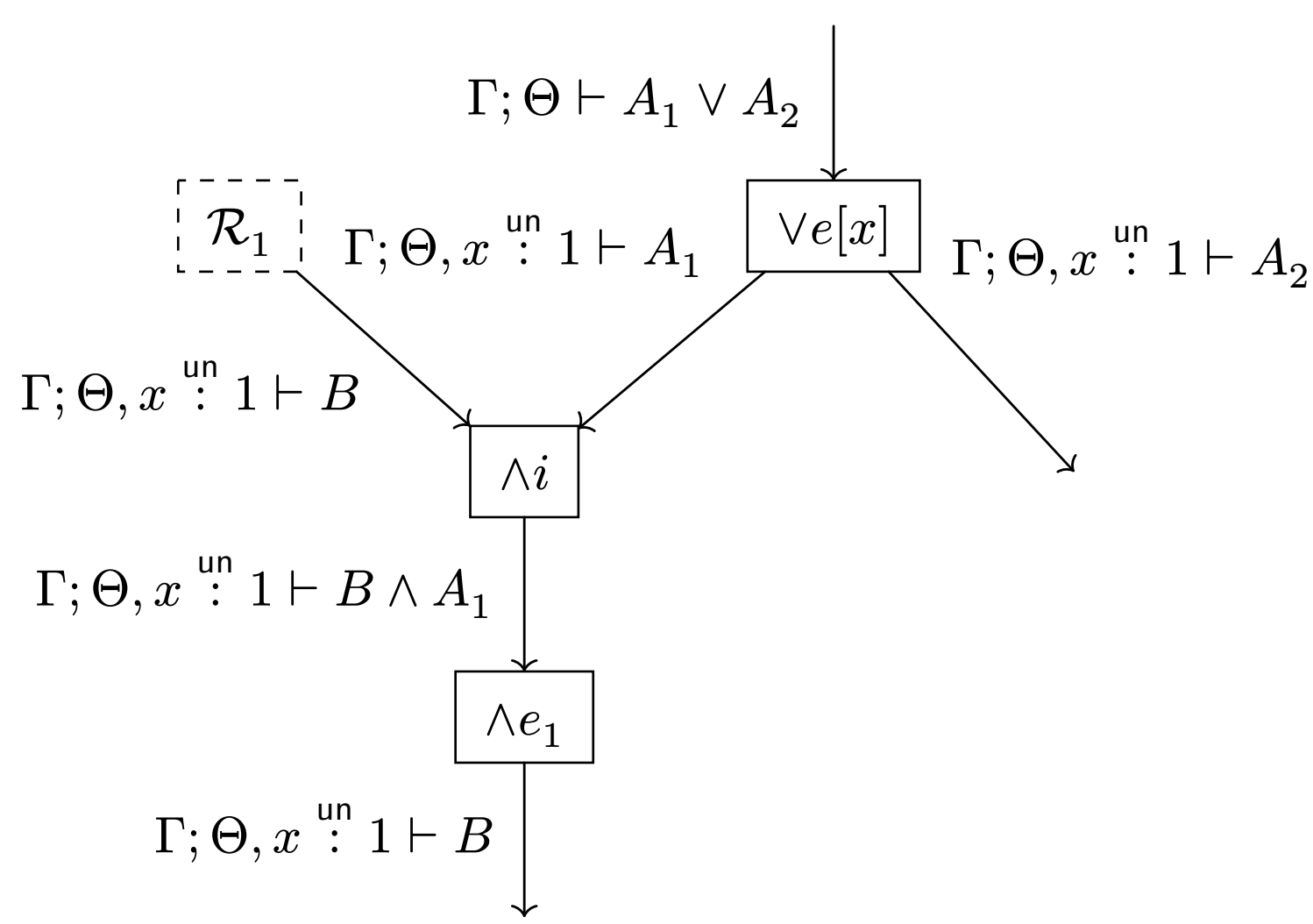

Figure 15: A subgraph of the left-hand side of Figure 9

Yet there is no way to weaken with a declaration annotated by $\mathsf{un}$; it cannot be weakened by a take node, either, as whether the branch is taken is yet unknown. Moreover, if the conjunction elimination is reduced to the first conjunct, the first branch of the disjunction elimination gets linked to a hole, breaking the link that propagates branching information downwards. If the first branch ends up being the taken branch, then the take node would disappear into the hole.

The third and final new node we introduce to resolve this issue is a *merge node* whose purpose is the retain the link that propagates an annotation. A merge node weakens its primary premise with a control declaration from the control context of its secondary premise; it therefore reduces when the secondary premise is a take node by linking the primary premise with a take node and discarding the secondary premise. It has no reduction rule for drop nodes in Figure 16, since dropped branches will eventually be consumed by a hole from below.

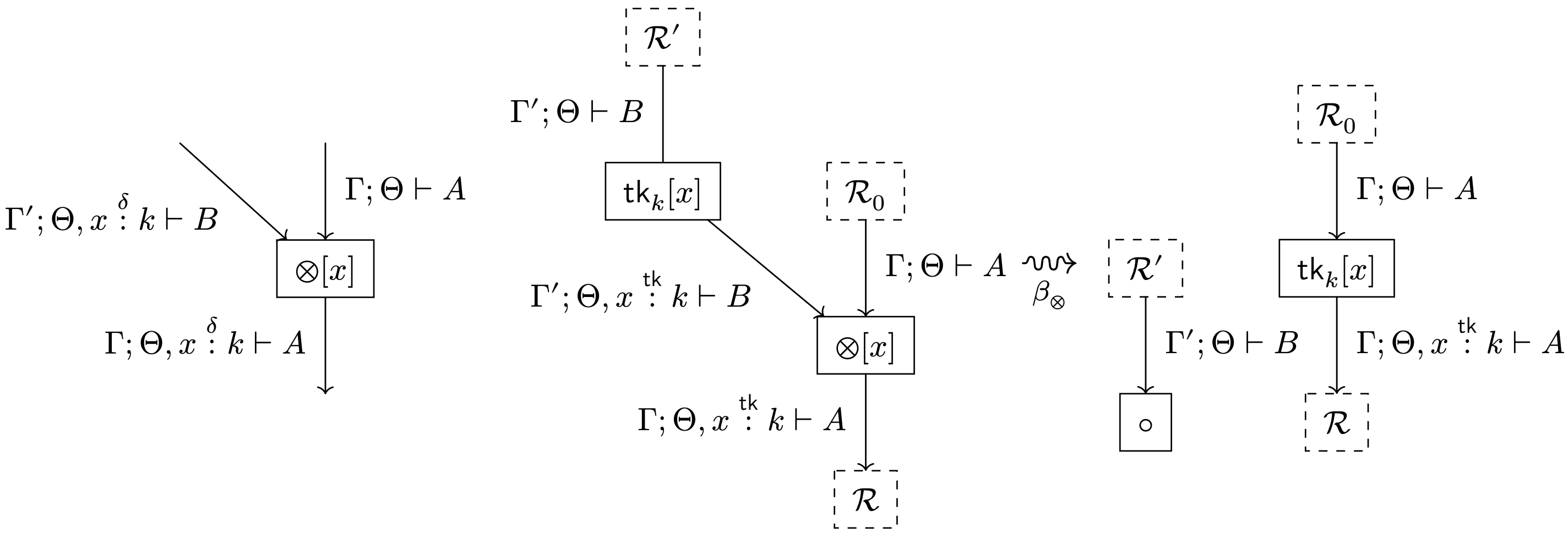

Figure 16: Typing and reduction rules for merge

Merge nodes then need to be inserted into the left-hand side of Figure 9 for the proof graph to be well typed. The updated graph is shown in Figure 17, additionally using forks to duplicate branches so that one duplicate propagates branching information to a merge node while the other is a conjunct as before. We leave as exercise to verify that this graph indeed reduces to the desired right-hand side along all reduction paths. In particular, if the conjunction eliminations are reduced first, then the merge node retains the link from the disjunction elimination to the join; if the disjunction elimination reduces first, then one of the merge nodes reduces when encountered by a take node. Additionally, each step preserves well-typedness, up to the un annotations updating to either a tk or dr annotation along the way.

At this point, a whopping six (!) new structural nodes have been introduced: joins, forks, and holes from the previous section, and now takes, drops, and merges to handle local reduction correctly. Three new annotations have also been added on top of control declarations: un, tk, and dr. The need for this machinery just to reduce disjunctions is rather unsatisfying, especially since take and drop nodes are transient nodes for reduction that one wouldn't write as part of a proof, and merge nodes have no real purpose other than to retain static information during typing. In the next subsection, we instead turn to a global reduction strategy, which avoids needing all of this new machinery.

### 3.2.2 Global Reduction

Permitting ourselves to examine the entire proof graph makes reduction more straightforward to express: when reducing a disjunction elimination, simultaneously find all join nodes and reduce them too. Then there is no longer any need to propagate branch information. Figure 18 illustrates this reduction rule for one join node. This is still a small-step reduction, despite looking at an arbitrarily large piece of graph.

When linking to and from taken branches, the control declaration is therefore removed from the context. To handle dropped branches, we introduce a *dead node* ▮, which doesn't remove the control declaration, but marks the branch as being dead. Retaining the control declaration allows that path to be consumed by a hole node, while marking the branch as dead prevents them from being used as premises to join nodes and from reïntroducing the pathological drop example from Figure 11. This is the only new node we need.

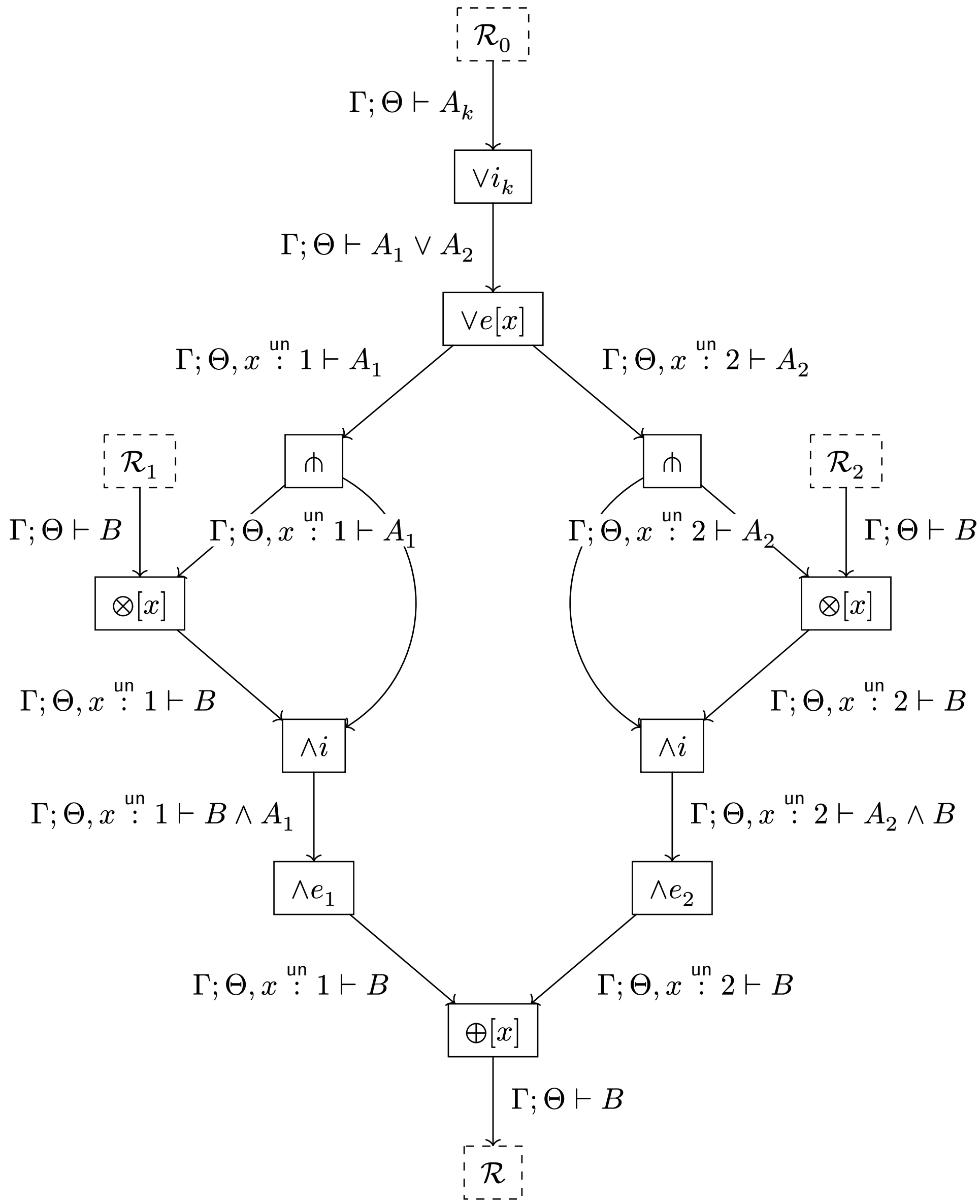

Figure 17: A well-typed, annotated variant of Figure 9 for local reduction

The entire proof graph is also examined globally during type checking to ensure that there is always exactly one $\vee e[x]$ node somewhere if there are $\oplus[x]$ nodes anywhere. Then the original pathological example from Figure 3 is ill typed because it's missing a corresponding disjunction elimination node that indicates which branch to take, not because leaf nodes may arbitrarily weaken the control context.

The proof graph is permitted to be disconnected. Such situations arise when a disjunction is used to dictate which branch is taken but its payload is useless, such as $\top \vee \top$ representing the booleans. For instance, in left-hand side of the example from Figure 9, if the conjunction eliminations in each branch are reduced first, the outgoing edges of the disjunction elimination gets linked to hole nodes, while the conjuncts get linked to the join node. Further reduction of the disjunction elimination followed by some hole node reductions yield the right-hand side as desired.

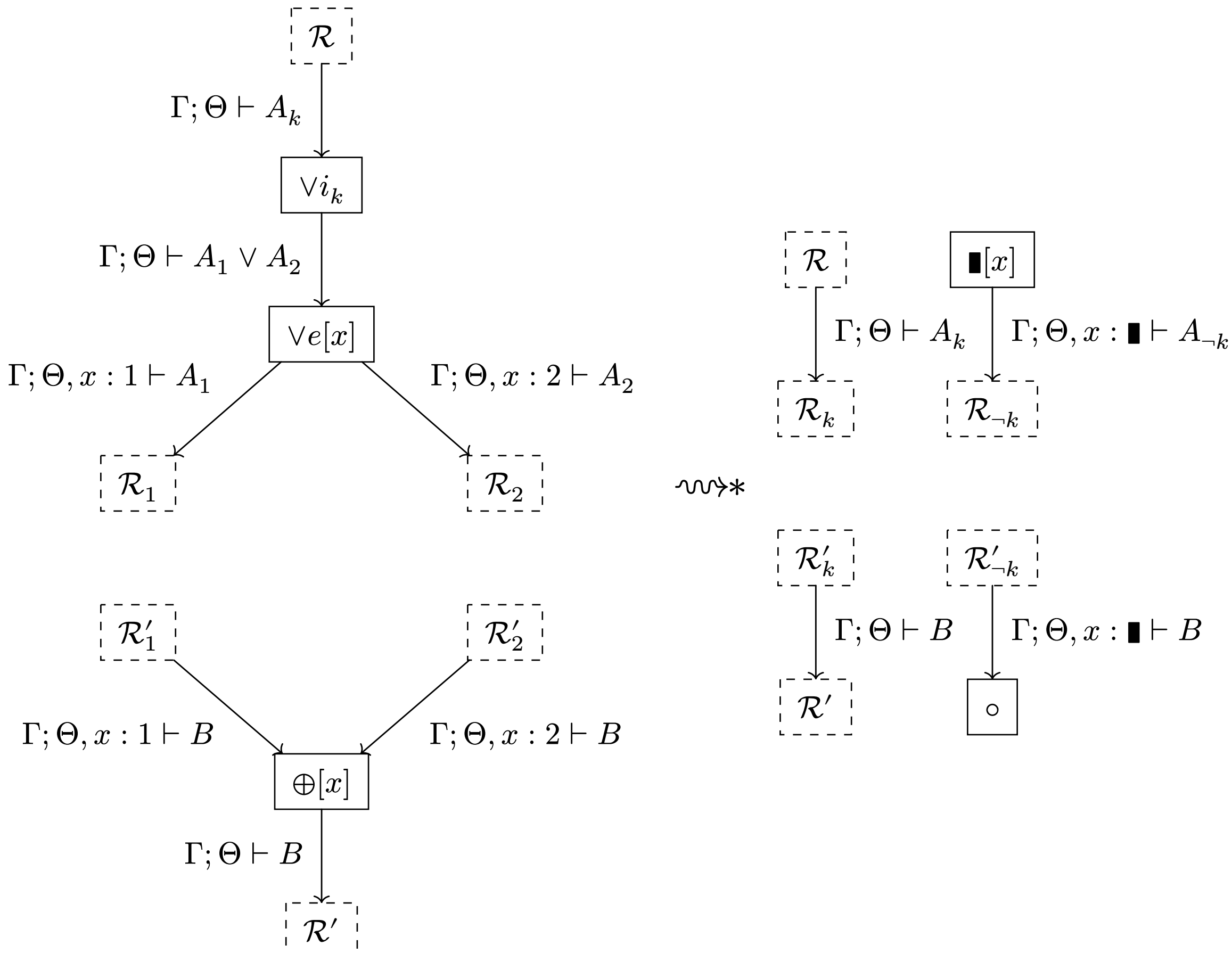

Figure 18: Global reduction rule for disjunction elimination and join

# 4 Commuting Conversions

Commuting conversions are reduction rules that permute (and sometimes duplicate) nodes surrounding structural nodes to unblock potential $\beta$-reductions. We have in fact already seen some: the reduction rules for take nodes in Figure 14 are commuting conversions. If a take node has an incoming edge from a conjunction introduction and an outgoing edge to a conjunction elimination, the take node commutes with the conjunction elimination, which allows the conjunction to reduce.

Generally, structural nodes need to commute with elimination nodes, moving the structural ones down and the elimination ones up. This corresponds to pushing elimination forms deeper into terms, such as in the case expression example from Section 1. (To distinguish commuting conversions from reduction rules in general, we use $\chi$ in place of $\beta$.)

$$(\text{case } cd \text{ of inl } c \Rightarrow f \mid \text{inr } d \Rightarrow g)\, a \underset{\chi}{\rightsquigarrow} \text{case } cd \text{ of inl } c \Rightarrow (f\, a) \mid \text{inr } d \Rightarrow (g\, a)$$

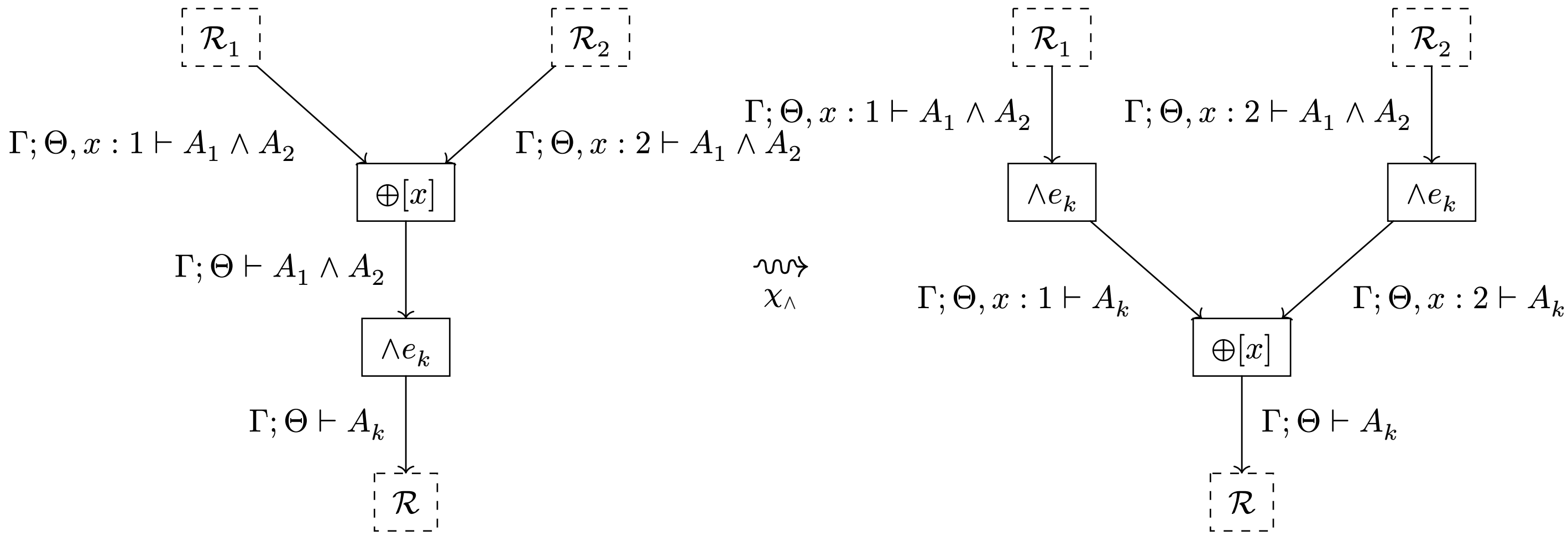

Figure 19: Commuting conversion for join and conjunction elimination

There are then three commuting conversions to consider: join nodes need to commute with conjunction elimination, implication elimination, and disjunction elimination. (There is no need to commute with false elimination as they don't reduce.) For local reduction of disjunction elimination, there would also be commuting conversions involving merge nodes, but they are convoluted and not particularly interesting, and we see shortly that there are problems with commuting conversions and local reduction even before reaching merge nodes.

Commuting conjunction elimination is fairly straightforward, since it only has one premise. Because join nodes have two premises, the conjunction elimination is duplicated within both branches after commutation in Figure 19.

Implication elimination only needs to commute with join nodes in the implication premise, not the argument premise, which may unblock a reduction if a join branch is an immediate implication introduction. This also duplicates the implication elimination, which then requires two uses of the argument, so it gets forked on the right-hand side of Figure 20.

On the right-hand side, $\mathcal{R}_3$'s control context is $\Theta$, but the new implication eliminations require their duplicate arguments' control contexts to be extended by $x : 1$ and $x : 2$, respectively. In the global reduction setting, the typing rule for fork nodes needs to be generalized so that they may also weaken the control context. (Correspondingly, the variable nodes' control contexts in Figure 5 may then be weakened by $\Theta_1, ..., \Theta_n$.) In the local reduction setting, additional merge and fork nodes are inserted to do this weakening.

The story of commuting disjunction elimination is more complicated. In the local reduction setting, we may only look locally at a single join node linked to a single disjunction elimination node. If we try to commute them and duplicate the disjunction elimination nodes across the branches, as illustrated in Figure 21, it becomes unclear how to combine the resulting four total branches back together. The $x$ branches cannot be joined immediately because the $y_1$ and $y_2$ branches haven't yet been handled. Even if join nodes were permitted to skip over other branches, we are still left with two different control variables $y_1, y_2$ rather than the one $y$ that $\mathcal{R}_3$ and $\mathcal{R}_4$ are expecting. It doesn't really make sense to join the $x$ branches first, either; this commutation is analogous to commuting a case-of-case expression,

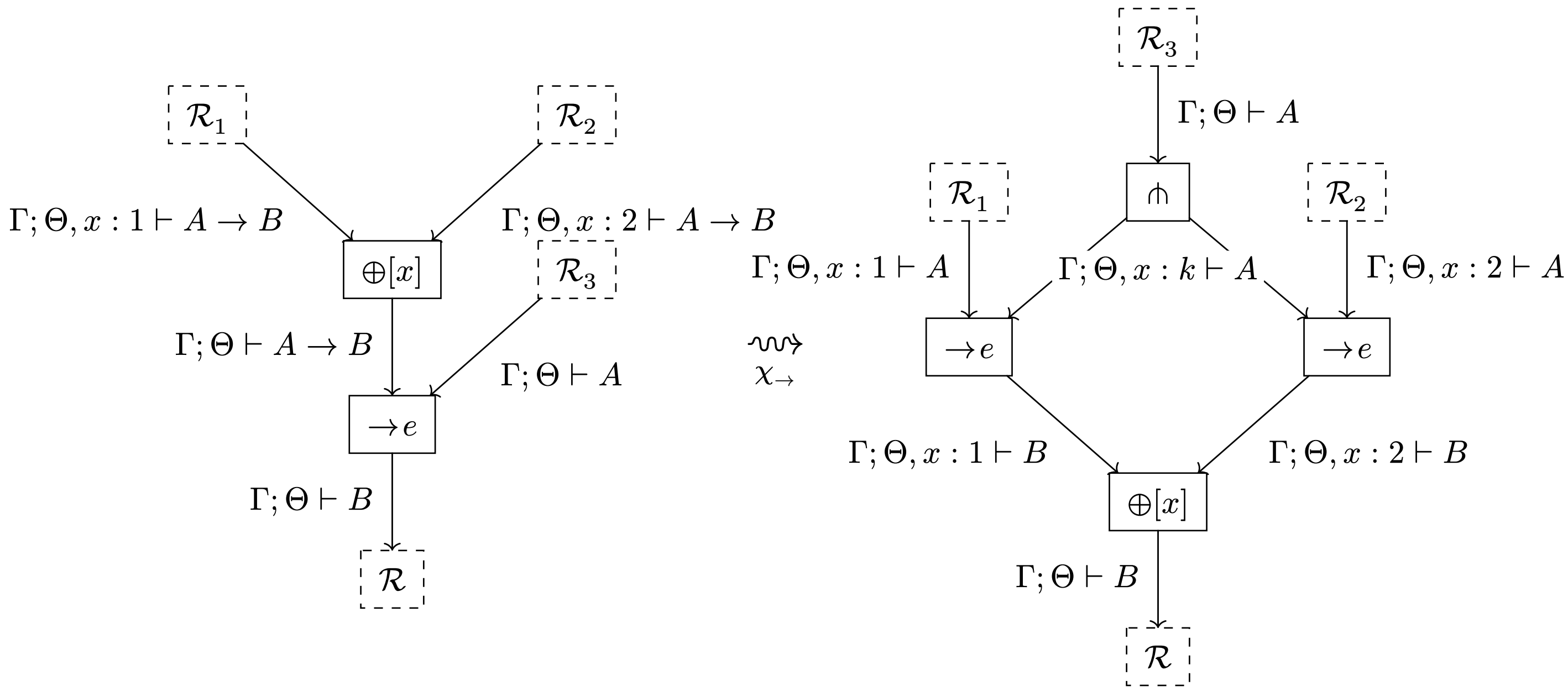

Figure 20: Commuting conversion for join and implication elimination

$$\begin{aligned} &\text{case (case } x \text{ of inl } u_1 \Rightarrow a \mid \text{inr } u_2 \Rightarrow b) \text{ of inl } v_1 \Rightarrow c \mid \text{inr } v_2 \Rightarrow d \\ &\rightsquigarrow \text{case } x \text{ of inl } u_1 \Rightarrow (\text{case } a \text{ of inl } v_1 \Rightarrow c \mid \text{inr } v_2 \Rightarrow d) \\ &\qquad\qquad \mid \text{inr } u_2 \Rightarrow (\text{case } b \text{ of inl } v_1 \Rightarrow c \mid \text{inr } v_2 \Rightarrow d), \end{aligned}$$

where the entire $x$ disjunction is moved outwards and doesn't get handled midway through the inner case expressions.

In the global reduction setting, commuting disjunction elimination more closely resembles commuting a case-of-case expression, since we may find an entire subgraph containing a disjunction elimination and its corresponding join and commute that subgraph wholesale, which is what

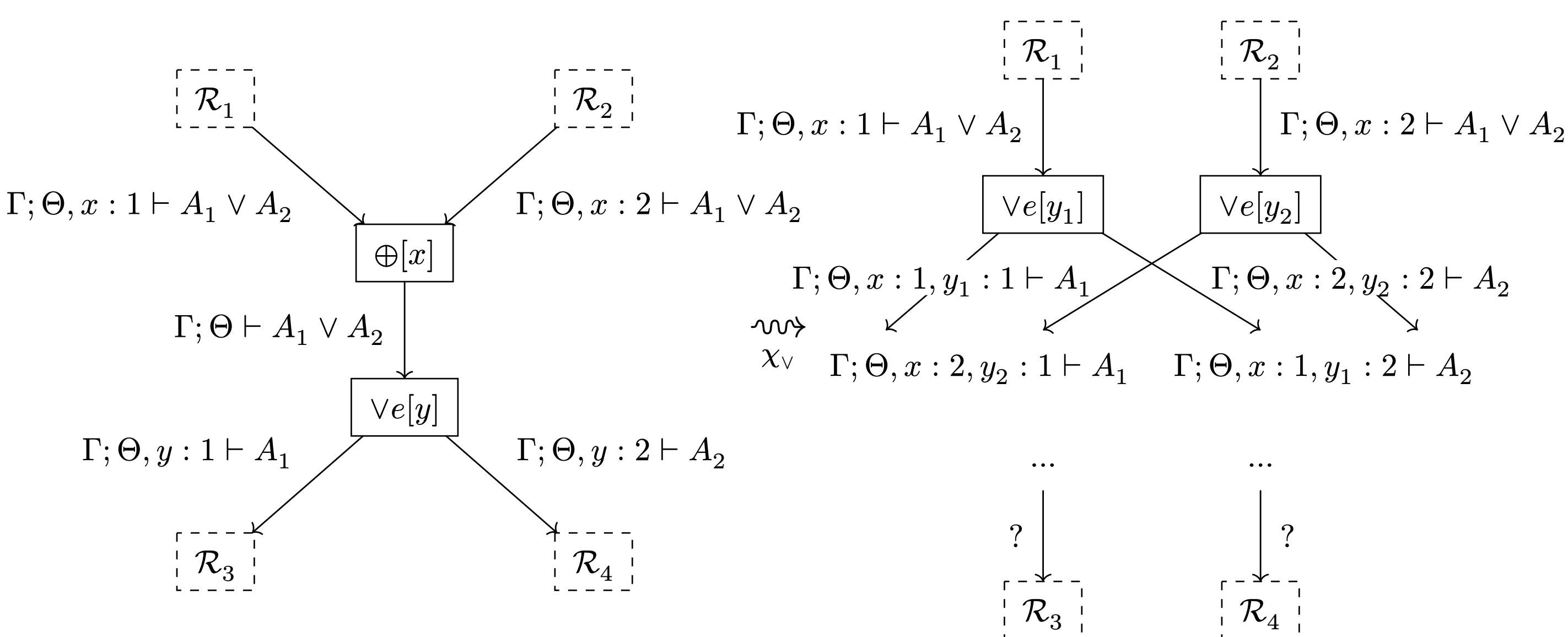

Figure 21: Local commutation rule for disjunction elimination (incomplete)

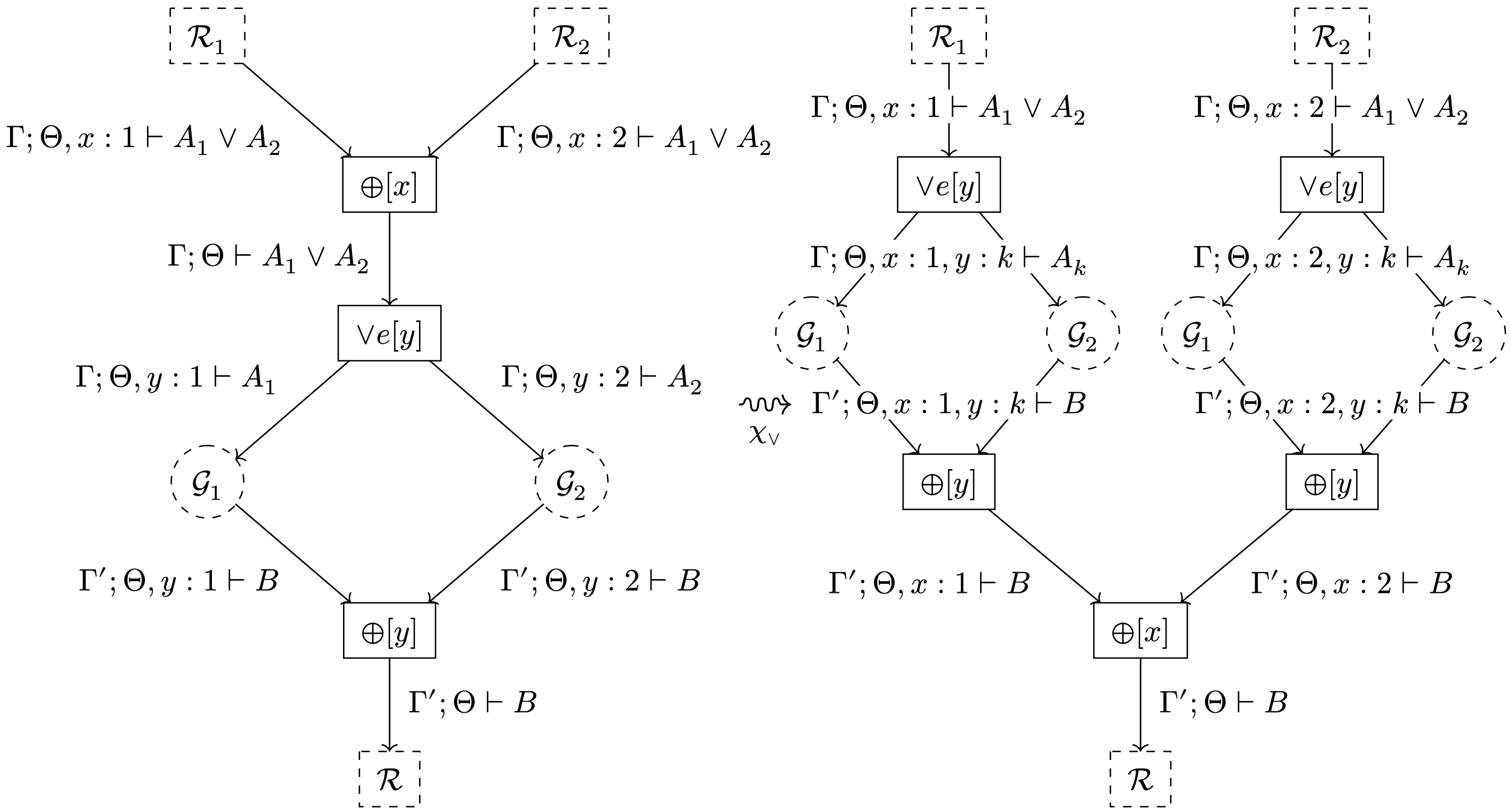

Figure 22: Global commutation rule for disjunction elimination

Figure 22 illustrates. Nevertheless, it remains unclear whether it would be possible to do so if the subgraph were disconnected between $\vee e[y]$ and $\oplus[y]$, or if there were multiple $\oplus[y]$ nodes.

# 5 Open Problems

In this work, we set out to design a proof language that accommodates Girard's disjunction elimination rule with multiple conclusions by using derivations in the form of DAGs, guided by the computational behaviour we expect from simplifying proofs. During this process, we introduce three structural nodes (join $\oplus[x]$, hole $\circ$, fork $\pitchfork$), and explore two ways of reducing disjunctions:

- A local strategy, whose reduction rules always examine a fixed number of nodes in a subgraph but require three additional structural nodes (take $\mathsf{tk}_k[x]$, drop $\mathsf{dr}_k[x]$, merge $\otimes[x]$); and
- A global strategy, which only requires one additional structural node (dead $\blacksquare[x]$), but involves typing and reduction rules that may examine an arbitrary number of nodes in the entire graph.

One of our goals was to have commuting conversions naturally arise as reduction rules for structural nodes. The results are incomplete: we have no solution for a local rule for commuting disjunction eliminations with join nodes, and only a specialized global rule for this commuting conversion. With the profusion of additional structural nodes and typing annotations, this may be the final nail in the coffin for a local reduction strategy, while a solution may lie in pushing the global reduction strategy further. Nevertheless, there are a number of outstanding technical issues that need handling:

1. Context weakening has been expressed throughout as extending the end of a context for simplicity, but for full generality, a fork node should be able to weaken in arbitrary locations in the context, *e.g.* from $\Gamma_1, \Gamma_2$ to $\Gamma_1, u : A, \Gamma_2$ and from $\Theta_1, \Theta_2$ to $\Theta_1, x : k, \Theta_2$, but with arbitrarily many variables. We don't anticipate any issues with this generalization, but expressing it formally is complicated. It would be easier to use de Bruijn indexing, where a multivariable weakening is a monotonic simultaneous renaming operation on contexts.

2. While the implication reduction rule in Figure 5 may be generalized to account for control context weakening, as was mentioned in Section 4, it's also possible for the control context of an ancestor of the implication introduction to have *fewer* declarations if that node is a disjunction elimination. When eliminating such an implication, the substituted argument would somehow need to be *strengthened* instead. Whether such a proof graph can actually be constructed and well typed is unclear.
3. We don't know yet whether well-typed proof DAGs remain acyclic after reduction. In particular, when reducing an implication in Figure 5, we don't know whether typing rules out paths from the function body subgraph $\mathcal{G}$ to the argument's root node $\mathcal{R}_0$. Because substitution replaces the variable nodes of $\mathcal{G}$ by links from $\mathcal{R}_0$, the existence of paths from $\mathcal{G}$ to $\mathcal{R}_0$ would yield a cycle. Since cyclic proof graphs might reduce perpetually, proving that reduction preserves acyclicity may be as difficult as proving strong normalization.
4. If we were to pursue the local reduction strategy, it must be noted that reducing disjunction eliminations changes the typing judgement on outgoing conclusions edges, since $\mathsf{un}$ annotations transform into $\mathsf{tk}$ or $\mathsf{dr}$ annotations. A type preservation lemma statement would need to account for this narrowing of the control context, although the top-level theorem on type preservation of incoming terminal edges would not, since their control contexts must be empty.

Once a proof graph language is designed with all issues resolved, the immediate future work is to formally prove type preservation of reduction. (Presumably it would use the global strategy.) For all reductions except for disjunction eliminations, it likely holds by inspection of their typing judgement edge labels.

For the disjunction-and-join reduction in Figure 18, proving type preservation requires showing that control declarations $x : k$ may be replaced by $x : \blacksquare$ or removed entirely without affecting well-typedness. Such a proof would likely require some form of induction, but as opposed to proof trees where induction hypotheses apply to subtrees, it seems unclear how an induction principle would work on subgraphs; this may require a foray into the graph rewriting literature.

Beyond type preservation, there are a number of natural properties and extensions to consider:

- Confluence (or the diamond property) of reduction, and (weak or strong) normalization of multistep reduction with respect to some normal proof graph form;
- Type-preserving compilation from proof trees to proof graphs to show that proof graphs haven't lost expressivity of natural deduction, and from proof graphs to proof trees to show that proof graphs aren't unexpectedly able to prove *more* propositions; and
- Extending from propositional logic to predicate logic with existentials, which may have alternate elimination forms just like disjunctions.

## A Grammar

$$
\begin{aligned}
u, v &::= \langle \text{typing context variables} \rangle \\
x, y &::= \langle \text{control context variables} \rangle \\
A, B, C, D &::= \top \mid \bot \mid A \to B \mid A \wedge B \mid A \vee B \\
\Gamma &::= \varnothing \mid \Gamma, u : A \qquad\qquad k ::= 1 \mid 2 \\
\Theta &::= \varepsilon \mid \Theta, x : k \mid \Theta, x \overset{\delta}{:} k \mid \Theta, x : \blacksquare \qquad \delta ::= \mathsf{un} \mid \mathsf{tk} \mid \mathsf{dr} \\
\mathcal{R} &::= \top i \mid \bot e \mid {\to} i[u] \mid {\to} e \mid \wedge i \mid \wedge e_k \mid \vee i_k \mid \vee e[x] \qquad \mathcal{G} ::= \langle \text{subgraph} \rangle \\
&\mid \bullet \mid \circ \mid \pitchfork \mid \oplus[x] \mid \otimes[x] \mid \mathsf{tk}_k[x] \mid \mathsf{dr}_k[x] \mid \blacksquare[x]
\end{aligned}
$$

## B Typing Rule Inventory

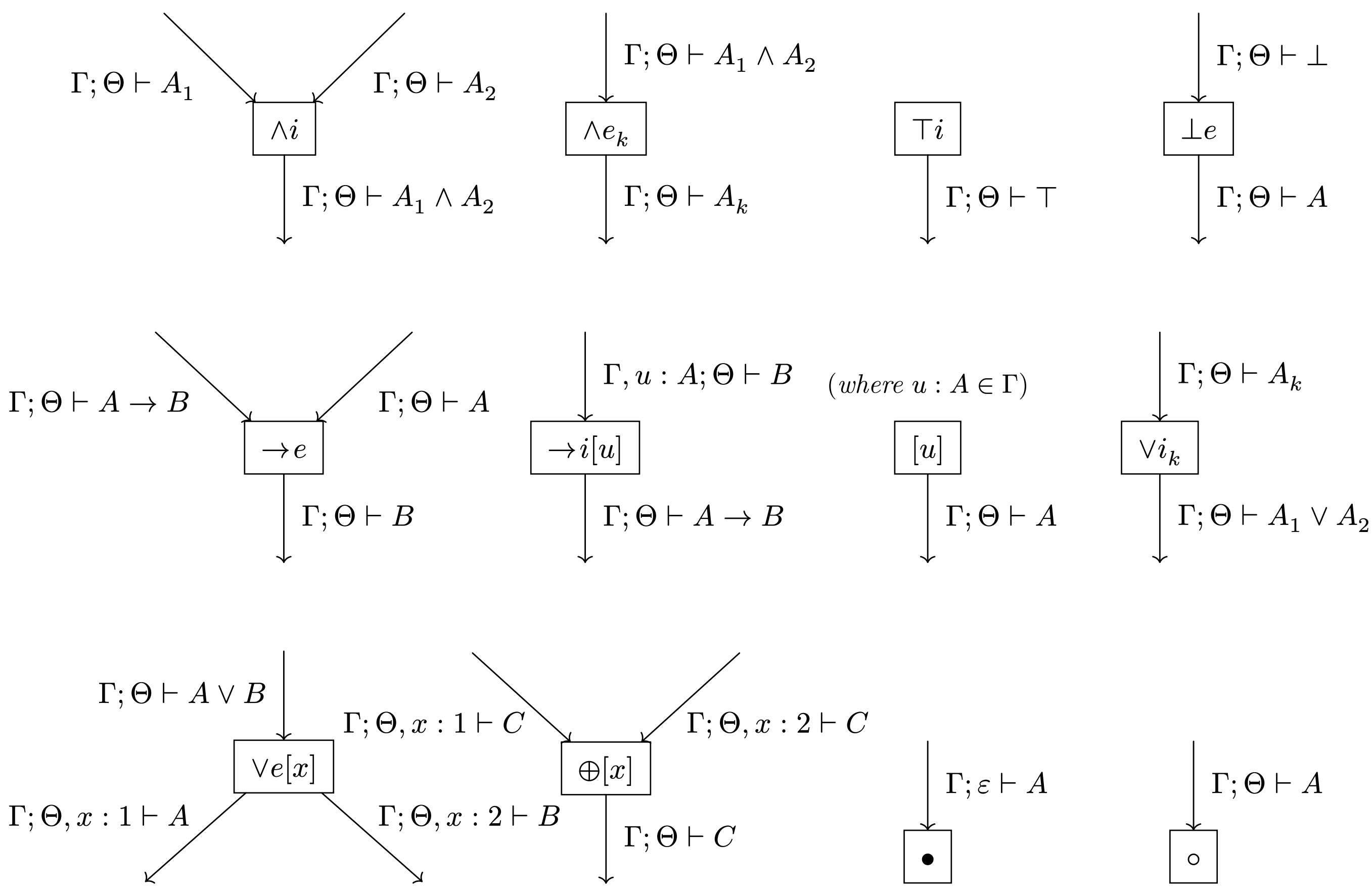

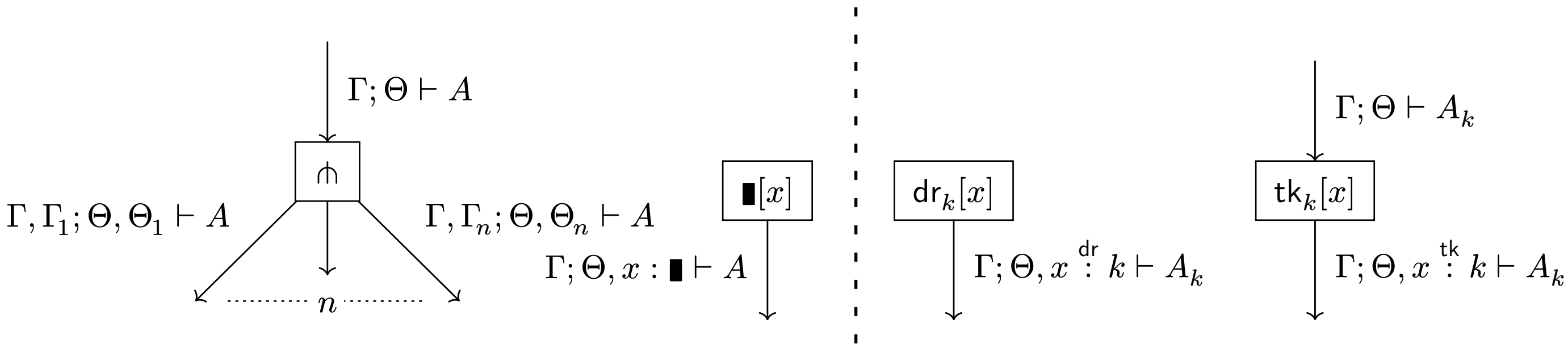
$\Gamma; \Theta \vdash A$
$\pitchfork$
$\Gamma, \Gamma_1; \Theta, \Theta_1 \vdash A$
$\Gamma, \Gamma_n; \Theta, \Theta_n \vdash A$
$n$
$\blacksquare[x]$
$\Gamma; \Theta, x : \blacksquare \vdash A$
$\mathsf{dr}_k[x]$
$\Gamma; \Theta, x \overset{\mathsf{dr}}{:} k \vdash A_k$
$\Gamma; \Theta \vdash A_k$
$\mathsf{tk}_k[x]$
$\Gamma; \Theta, x \overset{\mathsf{tk}}{:} k \vdash A_k$

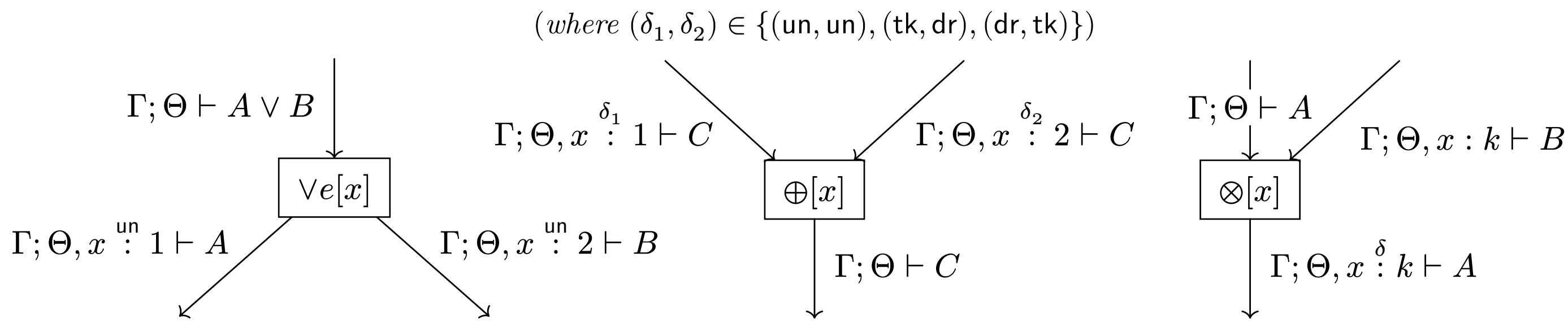
$(\textit{where } (\delta_1, \delta_2) \in \{(\mathsf{un}, \mathsf{un}), (\mathsf{tk}, \mathsf{dr}), (\mathsf{dr}, \mathsf{tk})\})$
$\Gamma; \Theta \vdash A \vee B$
$\vee e[x]$
$\Gamma; \Theta, x \overset{\mathsf{un}}{:} 1 \vdash A$
$\Gamma; \Theta, x \overset{\mathsf{un}}{:} 2 \vdash B$
$\Gamma; \Theta, x \overset{\delta_1}{:} 1 \vdash C$
$\Gamma; \Theta, x \overset{\delta_2}{:} 2 \vdash C$
$\oplus[x]$
$\Gamma; \Theta \vdash C$
$\Gamma; \Theta \vdash A$
$\Gamma; \Theta, x : k \vdash B$
$\otimes[x]$
$\Gamma; \Theta, x \overset{\delta}{:} k \vdash A$